# First-Order Phase Transition *vs*. Spin-State Quantum-Critical Scenarios in Strain-Tuned Epitaxial Cobaltite Thin Films

John E. Dewey[1], Vipul Chaturvedi[1], Tatiana A. Webb[2], Prachi Sharma[3], William M. Postiglione[1], Patrick Quarterman[4], Purnima P. Balakrishnan[4], Brian J. Kirby[4], Lucca Figari[1], Caroline Korostynski[1], Andrew Jacobson[1], Turan Birol[1], Rafael M. Fernandes[3], Abhay N. Pasupathy[2,5] and Chris Leighton[1]*

[1]*Department of Chemical Engineering and Materials Science, University of Minnesota, Minneapolis, Minnesota 55455, USA*

[2]*Department of Physics, Columbia University, New York, New York 10027, USA*

[3]*School of Physics and Astronomy, University of Minnesota, Minneapolis, Minnesota 55455, USA*

[4]*NIST Center for Neutron Research, National Institute of Standards and Technology, Gaithersburg, Maryland 60439, USA*

[5]*Condensed Matter Physics and Materials Science Division, Brookhaven National Laboratory, Upton, New York 11973, USA*

ABSTRACT: Pr-containing perovskite cobaltites exhibit unusual valence transitions, coupled to coincident structural, spin-state, and metal-insulator transitions. Heteroepitaxial strain was recently used to control these phenomena in the model $(Pr_{1-y}Y_y)_{1-x}Ca_xCoO_{3-\delta}$ system, stabilizing a nonmagnetic insulating phase under compression (with a *room-temperature* valence/spin-state/metal-insulator transition) and a ferromagnetic metallic phase under tension, thus exposing a potential spin-state quantum critical point. The latter has been proposed in cobaltites and can be probed in this system as a function of a disorder-free variable (strain). We study this here *via* thickness-dependent strain relaxation in compressive $SrLaAlO_4(001)/(Pr_{0.85}Y_{0.15})_{0.70}Ca_{0.30}CoO_{3-\delta}$ epitaxial thin films to quasi-continuously probe structural, electronic, and magnetic behaviors across the nonmagnetic-insulator/ferromagnetic-metal boundary. High-resolution X-ray diffraction, electronic transport, magnetometry, polarized neutron reflectometry, and temperature-dependent magnetic force microscopy provide a detailed picture, including abundant evidence of temperature- and strain-dependent *phase coexistence*. This indicates a first-order phase transition as opposed to spin-state quantum-critical behavior, which we discuss theoretically *via* a phenomenological Landau model for coupled spin-state and magnetic phase transitions.

*email: leighton@umn.edu

**INTRODUCTION**

Spin-state crossovers and transitions, where the spin of some ion gradually or abruptly changes *vs*. temperature, pressure, *etc*., are phenomena that have drawn interest in materials as diverse as metalorganic complexes[1–3], earth materials[4–6], and transition-metal oxides[7–13]. In oxides, $LaCoO_3$ is the archetypal spin-crossover compound[7–13]. At zero temperature (*T*), the $Co^{3+}$ ions in $LaCoO_3$ adopt a zero-spin $t_{2g}^6 e_g^0$ configuration, leading to a nominally diamagnetic ground state. At as low as 30 K, however, finite spin states turn on due to occupation of $e_g$ orbitals, leading to thermally-excited paramagnetism[7–13]. The details of this problem have proven challenging to understand, due to the delicate balance between crystal field splitting, Hund exchange, Co-O hybridization, and spin-orbit coupling[7–13].

More recently, a new dimension to the spin-state problem was uncovered in Pr-containing cobaltites such as $Pr_{0.5}Ca_{0.5}CoO_3$. The latter exhibits not a thermally excited spin-state crossover but a first-order spin-state *transition*, coupled to simultaneous structural and metal-insulator transitions[14–19]. On cooling below ~90 K, $Pr_{0.5}Ca_{0.5}CoO_3$ in fact exhibits a first-order spin-state transition, an unusual isomorphic structural transition (*Pnma* → *Pnma* with a ~2 % unit cell volume collapse), and a first-order metal-insulator transition (MIT), none of which were previously known in perovskite cobaltites[14,15]. Interest intensified when it was discovered that these phenomena occur due to a Pr valence shift. Specifically, various spectroscopies indicate that the typical $Pr^{3+}$ at room temperature abruptly shifts towards $Pr^{4+}$, decreasing the Co hole doping and triggering the spin-state and metal-insulator transitions[16–19]. The low-$T$/low-spin/low-cell-volume state was also found to be stabilized by substitution of smaller ions for $Pr$[20–25], $(Pr_{1-y}Y_y)_{1-x}Ca_xCoO_{3-\delta}$ emerging as a model system with valence transition temperature ($T_{vt}$) up to ~135 K at $y = 0.15$[25].

Perhaps the most intriguing aspects to the above are the connections that can be made with anomalous valence phenomena in other rare-earth (R) systems. Beyond the well-known anomalous valence tendencies of Eu[26], the most prominent example is the $RBa_2Cu_3O_{7-x}$ family, every member of which superconducts with the exception of R = Pr[27]. This is due to unusual Pr $4f$ - O $2p$ hybridization, leading to Fehrenbacher-Rice states near the Fermi energy[28]. Goodenough noted that Pr is the only R ion in perovskite and perovskite-related structures likely to have R $4f$ electrons near the Fermi energy and thus R-O hybridization and covalency[27]. Specifically in the $RCoO_3$ family, this was directly verified *via* density functional theory (DFT) calculations[29], which also reproduce the Pr valence transition in $Pr_{0.5}Ca_{0.5}CoO_3$[16]. In related $Pr_{0.5}Sr_{0.5}CoO_3$, Pr-O hybridization plays a key role in an unusual structural and magnetic transition[30]. More recently, Ramanathan *et al.* specifically targeted $Pr^{4+}$ compounds, seeking to exploit the unique features of Pr $4f^{\,1}$ - O hybridization for chemical design of novel quantum magnets[31]. Beyond perovskites, other R ions can play similar roles in different structures, one example being SmS, which exhibits a temperature-dependent valence/metal-insulator transition[32]. There are thus fascinating connections between R ion valence instabilities and transitions in multiple materials classes, but with little overarching understanding.

Using $(Pr_{1-y}Y_y)_{1-x}Ca_xCoO_{3-\delta}$ as an ideal system with which to elucidate the above, recent work from some of the current authors tuned the ground state not by chemical substitution in bulk but *via* heteroepitaxial strain in thin films[33]. Pseudomorphic epitaxy on various substrates was used to apply between -2.1 % (compressive) and 2.3 % (tensile) strain to $(Pr_{0.85}Y_{0.15})_{0.70}Ca_{0.30}CoO_{3-\delta}$, resulting in complete control over the electronic ground state[33]. Compressive strain was found to favor the low-spin/low-cell-volume state, driving $T_{vt}$ from ~135 K in bulk to 245 K under -2.1 % strain on $YAlO_3$ (YAO) substrates, confirmed by temperature-dependent scattering and spectroscopy[33]. Optimization of the composition to $(Pr_{0.75}Y_{0.25})_{0.70}Ca_{0.30}CoO_{3-\delta}$ then generated $T_{vt}$ = 291 K on YAO, realizing the first room-temperature valence transition in a perovskite oxide[33]. Conversely, tensile strain was found to quench the valence/structural/spin-state/metal-insulator transition, stabilizing a ferromagnetic (F) metal with Curie temperature ($T_C$) up to ~100 K[33]. This strain tuning thus generated "strain phase diagrams" of the type in Fig. 1(c) for the $(Pr_{0.85}Y_{0.15})_{0.70}Ca_{0.30}CoO_{3-\delta}$ composition, where $\varepsilon_{xx}$ is the "in-plane strain".

Such phase diagrams raise obvious questions, particularly the behavior in the region between the least compressive substrate ($SrLaAlO_4$ (SLAO), -0.63 % strain), with $T_{vt} \approx 140$ K and a nonmagnetic insulating ground state, and the least tensile substrate ($LaAlO_3$ (LAO), +0.25 % strain), with a F metallic ground state with $T_C \approx 50$ K. One intriguing possibility is a novel form of quantum critical point (QCP) at which the spin gap vanishes as a function of a non-thermal tuning parameter (strain in our case), triggering a transition from a non-magnetic to a F ground state[34]. This would contrast with better-established QCPs such as those between ordered magnetic states (*e.g.*, ferromagnetic and antiferromagnetic[35-37]), both of which obviously involve finite spin states[35-37]. Importantly, such a "spin-state quantum critical" scenario has been discussed in the context of perovskite cobaltites[34]. Tomiyasu *et al*. in fact probed this in bulk $LaCo_{1-x}Sc_xO_3$, claiming narrowing of the spin gap with increasing *x* due to increased covalency, and finding anomalous trends in magnetic, structural, and magneto-structural properties[34]. This was interpreted in terms of superposition of zero-spin and finite-spin states near the QCP[34]. Kuneš and Augustinský have also discussed condensation of spin-triplet excitons near the spin-state transition[38], including as an explanation for the behavior of $Pr_{0.5}Ca_{0.5}CoO_3$[39].

In light of the above, here we present a study that systematically probes the region between the long-range F metallic and nonmagnetic insulating phases in the $(Pr_{0.85}Y_{0.15})_{0.70}Ca_{0.30}CoO_{3-\delta}$ (hereafter "PYCCO") phase diagram (Fig. 1(c)). We achieve this *via* gradual relaxation of strain as a function of thickness (*t*), identifying the SLAO substrate (-0.63 % lattice mismatch) as optimal. A detailed picture of the structural relaxation is gained from *t*-dependent high-resolution X-ray diffraction (HRXRD), followed by extensive *T*-dependent transport, magnetometry, and magnetic force microscopy (MFM), as well as polarized neutron reflectometry (PNR) to confirm nominally depthwise-uniform magnetism. The results are unequivocal, indicating not suppression of the spin-state/metal-insulator transition followed by the onset of F metallic behavior as compressive strain is relaxed (*i.e.*, $T_{vt}$ vanishing prior to $T_C$ appearing), but instead phase coexistence over a substantial thickness (strain) range. At intermediate *t*, SLAO/PYCCO films in fact exhibit a clear $T_{vt}$ in resistivity on cooling, prior to a clear $T_C$ in magnetometry, with low-*T* magnetic phase coexistence confirmed by MFM. A phase diagram is thus mapped, revealing a significant phase coexistence region, meaning that a first-order phase transformation takes place as a function of strain, in contrast to the spin-state quantum-critical scenario. We discuss these findings theoretically *via* a phenomenological Landau model for the coupled spin-state and ferromagnetic transitions.

**RESULTS**

Fig. 1 first reviews the influence of heteroepitaxial strain on the electronic ground state of PYCCO films with $t \approx 11$ nm (30 unit cells (u.c.)), in the fully-strained, pseudomorphic limit[33] (see Methods section for synthesis details). Fig. 1(a) shows the $T$-dependent resistivity ($\rho$) of PYCCO films on YAlO$_3$(101) (YAO, $\varepsilon_{xx}$ = -2.1 %, red), SrLaAlO$_4$(001) (SLAO, $\varepsilon_{xx}$ = -0.6 %, blue), LaAlO$_3$(001) (LAO, $\varepsilon_{xx}$ = 0.3 %, green), and La$_{0.18}$Sr$_{0.82}$Al$_{0.59}$Ta$_{0.41}$O$_3$(001) (LSAT, $\varepsilon_{xx}$ = 2.3 %, black) substrates, along with bulk polycrystalline PYCCO for reference (black dotted line), where an MIT occurs below $T_{vt} \approx 135$ K. Under strong compression on YAO (red), a bulk-like MIT occurs, but with $T_{vt}$ promoted to ~245 K, the key result from our recent study[33]. Under weak compression on SLAO (blue), the MIT is broadened, potentially due to nanoscale doping inhomogeneity[33], but remains centered on $T_{vt} \approx 135$ K. The behavior under tension is very different: no $T$-dependent MIT occurs, leading to much lower $\rho$ at low $T$[33]. As in prior work, we refer to this behavior as "metallic", or "marginally metallic", despite the slightly negative $d\rho/dT$, due to the low $\rho$ and finite $\rho(T\rightarrow0)$[33]. Fig. 1(b) then shows the corresponding $T$-dependent magnetization ($M$), revealing that PYCCO films under compression are indeed non-F (due to the spin-state/metal-insulator transition on cooling), while films under tension are F, with $T_C$ rising from ~53 K on LAO to ~74 K on LSAT[33]. Recall that at this composition ((Pr$_{0.85}$Y$_{0.15}$)$_{0.7}$Ca$_{0.3}$CoO$_{3-\delta}$) no F metallic state exists in bulk[17,22–25], meaning that the F metallic behavior in tensile films in Fig. 1(b) is strain-stabilized[33].

As discussed in the Introduction, the results from Figs. 1(a,b) are summarized in Fig. 1(c), a "strain phase diagram", *i.e.*, a $T$-$\varepsilon_{xx}$ plot with $T_{vt}$ and $T_C$ labeled, where the white, green, and blue phase fields correspond to paramagnetic metal, nonmagnetic insulator, and long-range F metal, respectively[33]. This diagram includes low-$t$, fully-strained data on the aforementioned substrates (which are labeled at the top), plus SrLaGaO$_4$(001) (SLGO, $\varepsilon_{xx}$ = 1.7 %), as well as bulk behavior for reference. At $\varepsilon_{xx} > 0$ (tension), the dominant feature is the transition from paramagnetic metal to F metal on cooling below $T_C$. At $\varepsilon_{xx} < 0$ (compression), the dominant feature is the first-order spin-state/metal-insulator/structural/valence transition on cooling below $T_{vt}$, which is rapidly stabilized to higher $T$ with increasing compressive $\varepsilon_{xx}$. The resulting $T = 0$ evolution from nonmagnetic insulator to F metal as a function of $\varepsilon_{xx}$ is further illustrated in Fig. 1(d) by plotting $\rho$ (left axis, open circles) at 13 K (the lowest $T$ that could be measured in most cases) and $M$ at 10 K (right axis, closed circles). The latter was measured in a small 100 G (10 mT) in-plane field after field-cooling in 10 kG (1 T), and thus approximates the remnant magnetization ($M_r$). The ground state of tensile films is characterized by low $\rho$ (down to 3 mΩ cm on LAO) and $M_r \approx 0.2$-0.3 $\mu_B$/Co, reflective of the F metallic state. Under compression, however, $M_r = 0$ and the low-$T$ $\rho$ rises by over five orders of magnitude from LAO ($\varepsilon_{xx} = 0.3$ %) to SLAO ($\varepsilon_{xx} = -0.6$ %), becoming unmeasurable on YAO ($\varepsilon_{xx} = -2.1$ %). This reflects the rapid compressive-strain stabilization of the transition at $T_{vt}$ to the low-spin nonmagnetic insulating state.

The primary focus of the current work is to explore, using $t$-dependent strain relaxation in thicker films, what ground state(s) exist(s) in PYCCO films between $\varepsilon_{xx} = -0.6$ % and 0.3 %, *i.e.*, the unshaded low-$T$ region in Fig. 1(c). In order to pursue this, we first examine $t$-dependent strain relaxation on three candidate substrates – YAO, SLAO, and LAO – seeking continuous strain relaxation above some critical thickness ($t_{crit}$).

**Structural characterization of strain relaxation**

We focus first on PYCCO on SLAO, where the lattice mismatch is -0.6 % (blue points in Figs. 1(c,d)). Fig. 2(a) displays specular HRXRD scans around the PYCCO $002_{pc}$ (pc = pseudocubic) and SLAO 006 reflections of films with $t$ from 9 to 90 nm (labeled on the right). Consistent with prior work[33], at low $t$ (below ~50 nm), clear Laue fringes appear around the $002_{pc}$ film peaks, indicating low interface/surface roughness. As $t$ is increased, the fringe spacing decreases but then the fringes vanish at $t > 50$ nm, at which point the $002_{pc}$ film peak distinctly shifts towards the bulk PYCCO position (blue dashed line). Gaussian fitting of the $002_{pc}$ film peaks yields the $t$-dependence of the out-of-plane (OoP) lattice parameter ($c$) and Scherrer thickness normalized to the film thickness ($\Lambda/t$) in Figs. 2(b,c). For reference, Fig. 2(b) also shows the bulk pseudocubic lattice parameter ($c_{pc} = 3.780$ Å, blue dashed line), and the expected fully-strained OoP lattice parameter (red dashed line) assuming a Poisson ratio, $\nu$, between 1/4 and 1/3 (horizontal dashed lines), as common in cobaltites[40–43]. A clear decrease in $c$ occurs as $t$ is increased above ~50 nm (the additionally expanded lattice parameter in the very low $t$ limit could be associated with an increased density of defects such as O vacancies), resulting in films with $t = 90$ nm being over halfway relaxed. Correspondingly, $\Lambda/t$ abruptly decreases from ~1 (fully coherent) at $t < 50$ nm to much lower values at higher $t$, coincident with the loss of fringes in Fig. 2(a). This suggests the surface roughening and appearance/propagation of misfit dislocations that are common in conventional strain relaxation[44–48].

The strain relaxation of SLAO/PYCCO is further probed *via* the HRXRD rocking curves (RCs) around the film $002_{pc}$ peaks in Fig. 2(d). An initially narrow RC broadens monotonically with increasing $t$, resulting in RCs that can be fit (see Supplementary Information Fig. S1) to a sum of narrow Gaussian and broad Lorentzian peaks. Fig. 2(e) shows the resulting $t$ dependence of the full-width at half-maximum (FWHM) of the Gaussian (filled black circles) and Lorentzian components (open black circles), along with the ratio of Lorentzian to Gaussian intensities ($I_L/I_G$, red triangles, right axis). The broader Lorentzian is seen to dominate at high $t$, particularly above ~40 nm to 50 nm, increasing the total RC FWHM from 0.05° at $t = 12$ nm (Gaussian-dominated) to 0.165° at $t$ = 90 nm (Lorentzian-dominated). This is consistent with Figs. 2(b,c), suggesting strain relaxation, particularly above ~50 nm, inducing increases in roughness and dislocation density.

Definitive confirmation of strain relaxation is provided by Figs. 2(f,g), which are asymmetric reciprocal space maps (RSMs) around the SLAO 1011 and PYCCO $103_{pc}$ reflections for representative films with $t$ = 12 nm and 73 nm. At 12 nm (Fig. 2(f)), the film and substrate reflections occur at identical $Q_{x,y}$ (in-plane scattering wave vector), confirming identical in-plane (IP) lattice parameters, *i.e.*, pseudomorphic films. At 73 nm, however (Fig. 2(g)), the film $103_{pc}$ reflection becomes heavily streaked towards the expected bulk position (red ×), definitively confirming partial relaxation of IP and OoP lattice parameters, consistent with Figs. 2(a-e). Additional RSMs at other $t$ are provided in Supplementary Fig. S2, showing minor strain relaxation even at $t$ = 22 nm, prior to stronger relaxation above ~50 nm. The most direct way to quantify the extent of strain relaxation in these SLAO/PYCCO films is to use $\varepsilon_{xx}$, as in Figs. 1(c,d). As discussed in the caption to Fig. S2, however, accurate extraction of the $t$-dependent IP lattice parameters from RSM data was found challenging. Accurate determination of the OoP lattice parameter is simple, however (see Figs. 2(a,b)), and so for the remainder of this paper we quantify the strain relaxation using the average "OoP strain", $\langle\varepsilon_{zz}\rangle$. This is $\langle\varepsilon_{zz}\rangle = [c - c_{pc}]/c_{pc}$, where $c$ is the $t$-dependent $c$-axis lattice parameter (Fig. 2(b)), and $c_{pc}$ is the bulk pseudocubic lattice parameter.

With continuous strain relaxation of PYCCO on SLAO substrates established, we also probed strain relaxation on YAO (-2.1 % mismatch) and LAO (0.3 % mismatch). The equivalents of Fig. 2(a) for YAO/PYCCO and LAO/PYCCO are shown in Figs. S3(a,b). As recently reported[49], the strain relaxation of PYCCO on YAO is anomalous. YAO/PYCCO films in fact develop a bilayered structure above $t_{crit} \approx 30$ nm, where the bottom 20 nm to 30 nm remains fully strained (and misfit-dislocation-free), while the upper region almost fully strain relaxes, *via* a periodic misfit dislocation array pinned at a depth $\sim t_{crit}$[49]. This manifests in Fig. S3(a) as a distinct second film peak at higher $2\theta$ for $t > t_{crit}$, rapidly approaching the bulk PYCCO position. PYCCO films on LAO, on the other hand, do not relax to any detectable degree up to $t \approx 150$ nm, as shown in Fig. S3(b); this is not particularly surprising given the small lattice mismatch (0.3 %). Of the three substrates of potential interest for thickness-based tuning of PYCCO strain, LAO is thus not useful, while YAO induces *discontinuous* strain relaxation. We thus focus primarily here on SLAO/PYCCO, where the strain relaxation is conventional (Figs. 2, S1, S2), continuous *vs*. $t$ (Fig. 2(a-e)), and significant in magnitude (Fig. 2(b)). While the origin of the very different strain relaxation mechanisms on SLAO and YAO is not directly relevant here, it is likely associated with their different symmetries and octahedral tilt patterns[50].

**Impact of strain relaxation on electronic behavior**

Focusing next on the impact of strain relaxation on electronic properties, Fig. 3(a) shows $\rho(T)$ for SLAO/PYCCO films with $t$ from 12 nm to 90 nm. For comparison, $\rho(T)$ of bulk polycrystalline PYCCO is also shown (black dashed line), along with $\rho(T)$ of a thin, fully strained LAO/PYCCO film ($\varepsilon_{xx} = 0.3$ %, dark green dashed line). As explained in connection with Fig. 1(a), fully strained (*i.e.*, $\varepsilon_{xx} = -0.6$ %) SLAO/PYCCO films such as the $t = 12$ nm film in Fig. 3(a) exhibit similar insulating behavior to the bulk at low-$T$, but with a broadened $T$-dependent valence/spin-state/metal-insulator transition[33]. The $T_{vt}$ of such films can nevertheless be accurately determined from Zabrodskii analysis[51], as in Fig. 3(b). Plotted here is ln $W$ *vs*. ln $T$, where $W = -d\ln\rho/d\ln T$ is essentially a reduced activation energy[51], revealing the $T$-dependent transition as a clear peak. This peak has been definitively associated in prior work with the coupled valence/spin-state/metal-insulator transition, including via temperature dependent scattering and spectroscopic measurements[33,49]. As shown in Fig. S4 and discussed in the associated caption, the peaks in Fig. 3(b) can be fit to skewed Gaussians, yielding $T_{vt}$ (red dots in Fig. 3(b)), transition widths, *etc*.

Two key findings emerge from the $t$-dependent SLAO/PYCCO transport data in Figs. 3(a,b). First, the low-$T$ resistivity drops by more than four orders of magnitude as $t$ is increased from 12 nm to 90 nm (Fig. 3(a)), with $\rho(T)$ at $t = 90$ nm appearing to approach a finite intercept as $T \to 0$, *i.e.*, marginally metallic behavior. This is reinforced in Fig. 3(b), where high-$t$ films exhibit decreasing $W$ with decreasing $T$, another indication of metallicity as $T \to 0$[52]. A low-$T$ transition from insulating to weakly metallic is thus evidenced with increasing $t$. Second, $T_{vt}$ decreases as $t$ is increased (Fig. 3(b)), particularly for $t > (40$ to $50)$ nm, but remains far from $T_{vt} = 0$. The result at $t = 90$ nm is thus a PYCCO film that exhibits a distinct signature of a well-defined $T_{vt}$ (Fig. 3(b)) but nevertheless retains marginally metallic character as $T \to 0$ (Figs. 3(a,b)). The origin of this apparent dichotomy (a signature of a $T$-dependent MIT on cooling in samples that retain metallicity as $T \to 0$) will be clarified below. For reference, Figs. 3(a,b) also show the $\rho(T)$ and corresponding Zabrodskii plot for a fully strained ($\varepsilon_{xx} = 0.3$ %) LAO/PYCCO(10 nm) film. The F metallic nature of that tensile-strained film is apparent in Fig. 3(b), where $T_C$ appears as a broad hump. As the compressive strain is relaxed with increasing $t$ in SLAO/PYCCO films, $\rho(T)$ approaches that of LAO/PYCCO (Fig. 3(a)), but, again, $T_{vt}$ decreases slowly (Fig. 3(b)), certainly not approaching $T_{vt} = 0$ prior to the onset of metallicity.

The trends from Figs. 3(a,b) are summarized in Figs. 3(c,d), which plot the $t$ dependence of $T_{vt}$ and $\rho$(13 K) (blue circles, top axes). Fig 3(c) shows that $T_{vt}$ drops with increasing $t$, particularly above ~50 nm, but by only 18 K, from 135 K to 117 K. Fig. 3(d), however, shows a drop in $\rho$(13 K) of more than four orders, again with a noticeable slope change around 50 nm. The bottom axes of these plots then use the $c(t)$ from Fig. 2(b) to convert the dependences on $t$ to dependences on the OoP strain $\langle \varepsilon_{zz} \rangle$, which are shown as black circles (bottom axes). Strain relaxation (decreasing $\langle \varepsilon_{zz} \rangle$) leads to a large decrease in $\rho$ (13 K) (Fig. 3(d)), accompanied by only a modest decrease in $T_{vt}$ (Fig. 3(c)). Strain relaxation in SLAO/PYCCO films thus drives the system from a strongly insulating state with high $T_{vt}$ towards metallicity, but $T_{vt}$ remains apparent at all $t$.

Similar $t$-dependent transport measurements were also performed on LAO/PYCCO and YAO/PYCCO films. As expected based on Fig. S3(b), which established no strain relaxation on this substrate, LAO/PYCCO films exhibit $t$-independent $\rho(T)$ (Fig. S5(b)). In the YAO/PYCCO case, however, the discontinuous strain relaxation (Fig. S3(a) and Ref. [49]) leads to films with $t > t_{crit}$ exhibiting *two* $T_{vt}$ values (Fig. S5(a)): One from the fully strained bottom portion of the films ($T_{vt} \approx 245$ K) and one much lower $T_{vt}$ from the strain-relaxed upper portion. The result in Fig. S5(a) is a similar overall decrease in low-$T$ resistivity with increasing $t$ to that in Fig. 3(a), and a similar decrease in $T_{vt}$, but without the continuous tunability afforded by SLAO/PYCCO.

**Impact of strain relaxation on magnetic behavior**

Analogous to Fig. 3, Fig. 4 shows the $t$-dependent magnetic properties of SLAO/PYCCO films. Fig. 4(a) plots $M(T)$ in a 100 G (10 mT) in-plane field (essentially $M_r$), while Fig. 4(b) plots $dM/dT$, in both cases for $t$ from 10 nm to 90 nm; a LAO/PYCCO(21 nm) film is also shown for comparison. The fully strained SLAO/PYCCO(10 nm) and LAO/PYCCO(21 nm) films exhibit behaviors consistent with Fig. 1: F behavior with $T_C \approx 50$ K on LAO (tension), compared to no obvious F order on SLAO (compression). SLAO/PYCCO films at higher $t$, however, display progressively larger low-$T$ $M$, turning on at increasing $T$. The latter point is reinforced by Fig. 4(b), where $dM/dT$ reveals a relatively well-defined $T_C$, despite the absence of the clear order-parameter-like $M(T)$ on LAO (Fig. 4(a)). The emergence of marginally metallic behavior with increasing $t$ in Fig. 3 is thus accompanied by evidence of low-$T_C$ F behavior in Fig. 4. Figs. 4(c,d) summarize this, plotting the evolution of the apparent $T_C$ (estimated by the two tangent method) and $M(10$ K$)$ vs. both $t$ (blue circles, top axes) and $\langle \varepsilon_{zz} \rangle$ (black circles, bottom axes), analogous to Figs. 3(c,d). The apparent $T_C$ increases quickly with increasing $t$ (decreasing strain), accompanied by similar trends in $M(10$ K$)$. $T_C$ eventually reaches ~50 K, accompanied by $M_r \approx 0.05$ $\mu_B$/Co.

Particularly given the lack of order-parameter-like $M(T)$ for the strain relaxing SLAO/PYCCO films in Fig. 4(a), and their relatively low $M_r$, confirmation of long-range F order is important. This was achieved through PNR on a representative partially relaxed SLAO/PYCCO(22 nm) film, at $T = 5$ K in an in-plane field of 30 kG (3 T). Fig. 5(a) shows the non-spin-flip reflectivities $R^{++}$ and $R^{--}$ (where "+" and "-" indicate the neutron spin orientation of the incoming and outgoing beams relative to the sample magnetization) vs. specular scattering wave vector magnitude $Q_z$. The small but visible splitting of $R^{++}$ and $R^{--}$ immediately indicates long-range F order. This is emphasized in Fig. 5(b), which plots the $Q_z$ dependence of the spin asymmetry, SA = $(R^{++} - R^{--})/(R^{++} + R^{--})$, revealing clear oscillations indicative of a F thin film. A standard refinement (see Methods for fitting details and Table S1 for parameters) results in the solid lines through the data in Figs. 5(a,b), generating the chemical/nuclear and magnetic depth profiles in Fig. 5(c). Plotted in the latter are the depth ($z$) profiles of the nuclear scattering length density $\rho_{Nuc}$ (solid line, left axis) and $M$ (dashed line, right axis). $\rho_{Nuc}(z)$ is unsurprising, indicating a $\rho_{Nuc}$ for PYCCO within 2 % of bulk, and PYCCO surface roughness of approximately 3 to 4 unit cells (Table S1). More significantly, $M(z)$ reveals generally depth-wise-uniform magnetization, constant at $0.26 \pm 0.01$ $\mu_B$/Co through the majority of the film. The only deviation from this is associated with dead layer formation at the surface of approximately 2 to 3 unit cells (Table S1), comparable to other F cobaltite films[43,53,54]. The findings from Fig. 5(c) thus confirm long-range F in strain-relaxing SLAO/PYCCO films as well as the absence of significant property gradients through the film thickness. This is as expected for standard continuous strain relaxation, due to propagation of misfit dislocations[44–47]. Importantly, the $\langle\varepsilon_{zz}\rangle$ dependences of transport and magnetic properties in Figs. 3(c,d) and 4(c,d) can thus be simply and directly interpreted.

**Electronic phase coexistence in partially strain-relaxed films**

Fig. 6(a) combines the new information from Figs. 3, 4, and S5 (*i.e.*, the electronic and magnetic behavior during strain relaxation on SLAO and YAO) with the existing information in Fig. 1(d) (*i.e.*, from thin, fully strained films on all substrates). Plotted here is the $\langle\varepsilon_{zz}\rangle$ dependence of $M(10$ K) (closed circles, left axis) and $\rho(13$ K) (open circles, right axis) for PYCCO films on LSAT, SLGO, LAO, SLAO, and YAO, color coded as in Figs. 1(c,d). Multiple points on a particular substrate (*i.e.*, particular color) derive from different $t$, using strain relaxation to vary $\langle\varepsilon_{zz}\rangle$. The overall behavior is inverted from Fig. 1(d), due to the inverse relation between $\langle\varepsilon_{xx}\rangle$ and $\langle\varepsilon_{zz}\rangle$ through the Poisson ratio. Large negative $\langle\varepsilon_{zz}\rangle$ corresponds to large positive $\langle\varepsilon_{xx}\rangle$ (in-plane tension), which induces F order and thus increasingly large $M_r$. Conversely, large positive $\langle\varepsilon_{zz}\rangle$ corresponds to large negative $\langle\varepsilon_{xx}\rangle$ (in-plane compression), which induces high $T_{vt}$ and a strongly insulating nonmagnetic ground state. What is remarkable in Fig. 6(a) is the way the partially strain-relaxed PYCCO films on SLAO (blue points) connect the behavior between fully strained films on SLAO ($\langle\varepsilon_{zz}\rangle \approx 0.5$ %, with very low $M$ and very high $\rho$), and fully strained films on LAO ($\langle\varepsilon_{zz}\rangle \approx -0.2$ %, with significant $M$ and low $\rho$). There clearly exists a significant intermediate region ($-0.25$ % $< \langle\varepsilon_{zz}\rangle < 0.50$ %) in which non-zero F magnetization coexists with quite large resistivity. This arises because strain-relaxing SLAO/PYCCO films retain a well-defined $T_{vt}$ (Fig. 3(b,c)) even beyond the level of strain relaxation at which F order emerges (Fig. 4). This simultaneous signature of F magnetization and a $T_C$, along with a $T_{vt}$, should not arise in a single electronic phase in PYCCO; the spin-state transition to a nonmagnetic ground state below $T_{vt}$ is inconsistent with F ordering at $T_C$. Electronic phase coexistence is thus directly implicated, *i.e.*, spatial coexistence of F metallic and nonmagnetic insulating phases.

A vivid illustration is provided by Fig. 6(b), which plots $\rho(T)$ (black line, left axis) and $M(T)$ (red line, right axis) for an SLAO/PYCCO(73 nm) film, which is approximately halfway relaxed to bulk (Fig. 2(b)). $\rho(T)$ is quite flat down to ~150 K, below which it increases down to ~50 K, before flattening again. The Zabrodskii plot in the inset reveals that this is due to a well-defined $T_{vt}$ at ~117 K. This PYCCO film thus undergoes the signature of the valence/spin-state transition at ~117 K. Nevertheless, $M(T)$ still indicates the onset of F order at $T_C \approx 50$ K. Such data force us to conclude coexistence of F metallic and nonmagnetic insulating phases, as illustrated in the schematic at the top of Fig. 6(b). A uniform paramagnetic metallic (PM) state at high $T$ evolves below $T_{vt}$ into a spatially inhomogeneous distribution of PM (white) and nonmagnetic insulating (NI, green) regions. Further cooling below $T_C$ then transforms the PM regions to ferromagnetic metallic (FM, blue), resulting in a low-$T$ state consisting of coexisting FM and NI phases. At this specific thickness/strain, the FM volume fraction is apparently sufficient to a yield a marginally metallic (percolated) state with finite $\rho(T\rightarrow 0)$ (and ln$W$ decreasing with decreasing $T$ in the inset).

Seeking a direct test of this picture, $T$-dependent MFM was performed on a partially strain-relaxed SLAO/PYCCO(60 nm) film. Per Fig. 2(b), such films are approximately halfway relaxed at this $t$, resulting in $T_{vt} \approx 126$ K and $T_C \approx 45$ K (Figs. 3(a-c) and 4(a-c)), similar to the 73 nm film in Fig. 6(b). Strikingly, as shown in Fig. 7(a), after zero-field cooling (ZFC) to 5 K, MFM images indeed reveal clear nanoscopic magnetic inhomogeneity. These images are constant-height maps of the shift in the resonance frequency of the MFM cantilever and magnetic tip, per the schematic in Fig. 7(e) (see Methods for additional measurement details). Figs. 7(b,c) further demonstrate that this nanoscale magnetic inhomogeneity only becomes more apparent with increasing magnetic field up to 6 T (60 kG) (see Figs. S6(a-f) for images at additional field values), clearly never reaching the uniformly magnetized situation that would be expected in a phase-pure ferromagnet. Returning the field to zero (Fig. 7(d)) leaves the inhomogeneity largely unchanged. Further analysis *via* Fourier transforms (Fig. S6(g)) and autocorrelation analysis (Fig. S6(h,i)) confirms that the magnetic inhomogeneity in Figs. 7(a-d) occurs across broad length scales. Autocorrelation half-widths suggest characteristic scales centered on 40 nm to 70 nm, but broadly distributed.

Focusing on the origin of the inhomogeneity, Fig. 7(f) confirms that the magnetic inhomogeneity disappears on warming above $T_C$. Figs. S6(j-m) further demonstrate that the magnetic texture may have some correlation with topography but is not completely topography-driven. Fig. S7 also establishes that this inhomogeneity is not an artefact of work function variations. As established by Fig. S8, the MFM images in Fig. 7 are also very different to those that arise due to magnetic domains in phase-pure F cobaltite thin films. Fig. S8 shows the labyrinth domains[55–57] typical in LAO/La$_{0.5}$Sr$_{0.5}$CoO$_3$ films, for example, which are very different in character to the inhomogeneity in Fig. 7 and are annihilated in large fields, as expected. Moreover, Fig. 7(g) adds the vital observation that no such magnetic inhomogeneity arises in SLAO/PYCCO films with $t < t_{crit}$. The data in this case are from a SLAO/PYCCO(10.2 nm) film, which (see Fig. 2) is fully strained. Together, these observations thus establish that the nanoscale inhomogeneity apparent in Figs. 7(a-d) is definitively magnetic in origin, distinctly different to domain patterns in fully magnetized phase-pure F cobaltite films, and disappears in fully strained films on SLAO. This is strong confirmation of the schematic picture at the top of Fig. 6(b), the images in Fig. 7(a-d) essentially being direct images of the postulated FM/NI inhomogeneity. While on a scale of tens to hundreds of nm, the F regions under these conditions are apparently sufficiently long-ranged to generate a distinct Curie point in $M(T)$ (Figs. 4(a,b)) and a clear signature in PNR (Fig. 5).

**Final strain phase diagram**

The above findings are summarized in Fig. 8, which is essentially a revision of Fig. 1(c) where the low-strain region is now filled in *via t*-dependent strain relaxation. This is a "strain phase diagram" featuring $T_C$ (solid symbols) and $T_{vt}$ (open symbols), as well as PM (white), FM (blue), NI (green), and mixed-phase regions. For reasons explained above, this is plotted *vs.* $\langle\varepsilon_{zz}\rangle$ rather than $\langle\varepsilon_{xx}\rangle$, inverting it with respect to Figs. 1(c,d). The key finding is that the $T_C$ of the FM phase field is clearly not suppressed to zero prior to the onset of the $T_{vt}$ associated with the NI phase field. In contrast, there is a substantial interval of strain (at least -0.1 % < $\langle\varepsilon_{zz}\rangle$ < 0.5 %) over which $T_C$ and $T_{vt}$ are both detected, and FM and NI phases coexist, in correspondence with Fig. 6(a). In addition to the points in Fig. 8 that derive from continuous *t*-dependent strain relaxation on SLAO (blue), we also plot here strain-relaxed points from YAO substrates (red). Despite the discontinuous relaxation on YAO[49], these points nevertheless overlap with all others. This can be seen in the $T_{vt}(\langle\varepsilon_{zz}\rangle)$ data, where the blue points (SLAO) and red points (YAO) collapse to a single curve. In addition, a very thick (150 nm) film on YAO with an $\langle\varepsilon_{zz}\rangle$ that almost perfectly vanishes is shown in Fig. 8, revealing a ferromagnetic $T_C \approx 58$ K (see Fig. S9 for additional details), nicely connecting the $t$-dependent $T_C$ of the films on SLAO (blue solid points) with the $t$-independent $T_C$ of the films on LAO (green solid points). While the strain in PYCCO films on YAO is undoubtedly inhomogeneous through the depth[49], the strain-relaxed upper film portions thus nevertheless have $T_{vt}$ and $T_C$ consistent with other results. The electronic/magnetic phase coexistence around $\langle \varepsilon_{zz} \rangle \approx 0$ in Fig. 8 is thus robust across all substrates.

We make two points regarding Fig. 8 before moving to a theoretical discussion. First, Fig. 8 bears a notable resemblance to a recently published $T$-$y$ phase diagram of bulk $(Pr_{1-y}Sm_y)_{0.7}Ca_{0.3}CoO_3$[58]. In this system, the slightly smaller radius of $Sm^{3+}$ *cf*. $Pr^{3+}$ enables tuning across the nominal FM/NI boundary, albeit compositionally rather than *via* strain. The authors of Ref. [58] also conclude the coexistence of $T_{vt}$ and $T_C$, short-range FM clusters featuring in their interpretation[58], in agreement with the current work. Second, while the right side of the phase coexistence region in Fig. 8 is well pinned down by the $t$-dependent data on SLAO/PYCCO, the left side is less definitive. In fact, comparing the MFM data of Fig. S7 (on LAO) with the data of Figs. S6 (on SLAO) and S9, suggests that the LAO/PYCCO films in Fig. 8 (green points) may not have entirely phase-pure FM. The left edge of the phase coexistence region may thus extend further to the left in Fig. 8, potentially symmetrizing the phase coexistence region with respect to $\langle \varepsilon_{zz} \rangle = 0$.

**DISCUSSION**

To shed light on the above, we employ a phenomenological Landau model for coupled spin-state and magnetic phase transitions. The spin-state transition is described in terms of an order parameter $\varphi$, where $\varphi > 0$ and $\varphi < 0$ correspond to high-spin (HS) and low-spin (LS) states, respectively, mapping to the valence-averaged high-spin/high-volume and low-spin/low-volume states in systems such as $(Pr_{1-y}Y_y)_{1-x}Ca_xCoO_{3-\delta}$. The free-energy near the spin-state transition, to fourth order, is:

$$F_{st}(\varphi) = \frac{1}{2}a(T,\varepsilon)\varphi^2 + \frac{1}{2}\varphi^4 - h(T,\varepsilon)\varphi. \quad (1)$$

Because $\varphi$ does not break any symmetries of the system, all powers of $\varphi$ are allowed in the free-energy expansion. We can omit the third-order term *via* an appropriate shift of the order parameter, however, and the prefactor of the quartic term can be set to unity through appropriate rescaling of order parameters. The Landau coefficients here, $a$ and $h$, depend on temperature $T$ and strain $\varepsilon$. Assuming a simple analytic dependence, we can rewrite them as $a(T,\varepsilon) = \alpha_1 \tilde{a}(T) + \beta_1 \tilde{h}(\varepsilon)$ and $h(T,\varepsilon) = \alpha_2 \tilde{a}(T) + \beta_2 \tilde{h}(\varepsilon)$, where $\alpha_i$ and $\beta_i$ are constants, and the functions $\tilde{a}(T)$ and $\tilde{h}(\varepsilon)$ depend only on temperature and strain, respectively, *via* the standard relationships $\tilde{a}(T) = \tilde{a}_0(T - T_{st})$ and $\tilde{h}(\varepsilon) = \tilde{h}_1(\varepsilon - \varepsilon_{st})$, where $T_{st}$, and $\varepsilon_{st}$ denote the temperature and strain values at the spin-state transition. The parameter space of $(\tilde{a}, \tilde{h})$ can thus be understood as the $(T,\varepsilon)$ parameter space in the phase diagram of Fig. 8. Hereafter, we omit any functional dependence on $T$ and $\varepsilon$ unless explicitly required for clarity, presenting our results in the $(\tilde{a}, \tilde{h})$ space.

Minimizing the free energy with respect to $\varphi$ gives the phase diagram in Fig. 9(a) in the $(\tilde{a}, \tilde{h})$ parameter space, where we chose $\alpha_1 = 4$, $\beta_1 = 0.5$, $\alpha_2 = 5$, $\beta_2 = -2$. Unsurprisingly, we obtain the characteristic phase diagram of the liquid-gas transition of water, with a first-order transition line at $a < 0$ ending at the critical endpoint $a = 0$, $h = 0$, corresponding to $T = T_{st}$, and $\varepsilon = \varepsilon_{st}$. For $a > 0$, the free energy exhibits a single minimum corresponding to either the HS (yellow in Fig. 9(a)) or LS (green in Fig. 9(a)) state depending upon the sign of $h$. In this region of the phase diagram, there is no phase transition between these two states, but instead a smooth crossover, marked by the black dotted line. For $a < 0$ in Fig. 9(a), instead of the first-order transition line, we show the upper and lower spinodal lines (dotted red), which correspond to the limit of metastability of each phase. In the region between these spinodal lines (lighter-green), the free energy has two local minima, implying that the system is in a regime of HS-LS phase coexistence. Outside of this region, the free energy again has a single minimum, corresponding to HS or LS.

To understand the F order observed in conjunction with the spin-state transition in our strained PYCCO system, we now introduce a F order parameter $m$. Because $m$ breaks time-reversal symmetry, it can couple to the spin-state order parameter $\varphi$ only through a linear-quadratic coupling with coefficient $\lambda$, which we choose to be positive, *i.e.*, $\lambda > 0$. The free energy in the vicinity of this coupled spin-state and ferromagnetic transition is given by:

$$F(\varphi, m) = \frac{1}{2} a\, \varphi^2 + \frac{1}{4} \varphi^4 - h\, \varphi + \frac{1}{2} A\, m^2 + \frac{1}{4} m^4 - \lambda\, \varphi\, m^2. \quad (2)$$

Note, again, that the coefficients of the quartic terms are set to unity, which can always be done through appropriate rescaling of order parameters. The coefficient $A(\varepsilon, T)$ implicitly contains the strain-dependence of the bare Curie temperature $T_{C,0}(\varepsilon)$ (*i.e.*, with no coupling to the spin-state transition) *via* $A(\varepsilon, T_{C,0}) = 0$. In our model, for simplicity, we assume an essentially strain-independent $T_{C,0}$, and write $A(\varepsilon, T) = \tilde{A}_0 (T - T_{C,0})$. The actual Curie temperature $T_C(\varepsilon)$ depends on the spin-state of the system, and is given by:

$$T_C(\varepsilon) = T_{C,0} + \frac{2\lambda}{\tilde{A}_0} \varphi(\varepsilon), \quad (3)$$

where $\varphi(\varepsilon)$ is the solution that minimizes the free energy in Eq. (2). Thus, $\varphi > 0$ (the HS state) favors F order, whereas $\varphi < 0$ (the LS state) suppresses it. In particular, for a sufficiently large (in magnitude) LS order parameter $\varphi < -\frac{T_{C,0}\,\tilde{A}_0}{2\lambda}$, the F state is completely destroyed. Recall that our model, and thus the parametrizations of the Landau coefficients, are only valid in the vicinity of the joint spin-state and F transitions. As such, when comparing our results with experiment, it is important to bear in mind that the model cannot predict the fate of the spin-state and ferromagnetic transitions away from the multicritical point.

We can now obtain the phase diagram of the free energy of Eq. (2) in the $(\tilde{a}, \tilde{h})$ parameter space. To simplify the analysis, we assume $\tilde{A}_0 = \tilde{a}_0$, which allows us to re-write $A(\varepsilon, T) = \tilde{a}(T) + \Delta t$, where $\Delta t = \tilde{a}_0(T_{st} - T_{C,0})$ is proportional to the difference between the spin-state transition and bare Curie temperatures. Minimizing the free energy (2) with respect to both $\varphi$ and $m$, we obtain:

$$m = \begin{cases} \pm\sqrt{-(A - 2\lambda\varphi)}, & (A - 2\lambda\varphi) < 0 \\ 0, & (A - 2\lambda\varphi) > 0 \end{cases}, \quad (4)$$

and

$$a\,\varphi + \varphi^3 = h + \lambda(A - 2\lambda\varphi)\Theta(2\lambda\varphi - A), \quad (5)$$

where $\Theta(x)$ is the Heaviside step function, which is 1 for $x > 0$ and 0 for $x < 0$. Clearly, if $\lambda = 0$, we obtain independent spin-state and F orders, as shown in the phase diagram of Fig. 9(b).

For $\lambda > 0$, however, the phase diagram changes considerably, since non-zero $\varphi$ impacts the F transition *via* Eq. (3) and non-zero magnetization has a feedback effect on $\varphi$ *via* Eq. (5). The outcome, shown in Fig. 9(c), is the emergence of a ferromagnetic high-spin state (FHS) and the absence of a ferromagnetic low-spin state, since the magnetization is completely suppressed inside the low-spin phase (near the multi-critical point). Note that while it is formally possible for a ferromagnetic LS state to be stabilized, this only happens far from the multi-critical point for large $\lambda$, where our Landau expansion model is likely inapplicable. Importantly, the HS+LS coexistence region of Fig. 9(a) is split in Fig. 9(c) into a narrow region with spin-state coexistence and no F order, located immediately below the spin-state critical endpoint, and a wider region of FHS+LS coexistence. Consequently, a coupled first-order transition between the ferromagnetic HS state and the non-magnetic LS state occurs (provided we are not too close to the spin-state critical endpoint), with a coexistence region between the two phases emerging in between. This is further illustrated in Fig. 10, where we present the order parameters $\varphi$ and $m$ vs. $\tilde{h}$ at fixed $\tilde{a}$. The key point is that the phase diagram of Fig. 9(c) therefore qualitatively resembles the experimental phase diagram of Fig. 8 if we associate the metallic and insulating states with HS and LS states. This suggests that the temperature-strain phase diagram of PYCCO (Fig. 8) is well described by the phenomenological model proposed here. In this regard, note that the general structure of the phase diagram, and particularly the coexistence of FHS and LS phases, is not unique to the model parameters chosen here, but is a more general feature of the problem for sufficiently large coupling constant $\lambda$.

It is important to emphasize the limitations of this analysis. First, the phenomenological model is valid only in the vicinity of the spin-state critical endpoint, and, second, it cannot say anything about the microscopic mechanisms involved in the coupled transitions. Moreover, it tacitly assumes that the valence transition triggers the spin-state and metal-insulator transitions, and it is possible that the valence transition has a non-trivial relationship with the spin-state transition. Nevertheless, the above analysis indicates that a spin-state QCP could still potentially be realized in this system, provided that the spin-state critical endpoint could be tuned to $T = 0$. The most promising strategy would be to tune the starting $T_{vt}$, which is highly sensitive to composition[15,20,21,25]. In $(Pr_{1-y}Y_y)_{1-x}Ca_xCoO_{3-\delta}$, for example, $T_{vt}$ in bulk is controlled by $x$ and $y$[20,24,25], and evolves with the substitution of other R elements for Pr[21,58,59]. These approaches could be used to shift the position of 4$f$ states with respect to the Fermi level, thereby identifying compositions in which strain-based control of the type explored here could shift the spin-state critical endpoint towards $T = 0$. Future work in this direction would clearly be worthwhile.

**SUMMARY**

The PYCCO system is a model one amongst perovskite cobaltites, exhibiting coupled spin-state/structural/metal-insulator transitions driven by a fascinating valence shift mechanism of broad interest. Recent work establishing the "strain phase diagram" of epitaxial PYCCO thin films demonstrated complete control over the electronic/magnetic ground state, exposing a potential spin-state quantum critical point between nonmagnetic insulating and ferromagnetic metallic phases. Here, this potential quantum critical point has been comprehensively probed *via* quasi-continuous thickness-driven strain relaxation, assessed *via* high-resolution X-ray diffraction characterization, and coupled to electronic, magnetic, neutron reflectometry, and magnetic force microscopy measurements. The results are unequivocal, revealing not a spin-state quantum critical point, but instead spatial coexistence of nonmagnetic insulating and ferromagnetic metallic phases, evidencing a first-order phase transition *vs*. strain. This has been reconciled with theoretical expectations using a phenomenological Landau model for coupled spin-state and ferromagnetic transitions, leading to suggested routes to further explore a spin-state quantum critical point in systems like this.

**METHODS**

Polycrystalline $(Pr_{0.85}Y_{0.15})_{0.7}Ca_{0.3}CoO_{3-\delta}$ sputtering targets (2" diameter) were synthesized by solid-state reaction, cold pressing, and sintering of stoichiometric mixtures of $Pr_6O_{11}$, $Y_2O_3$, $CaCO_3$, and $Co_3O_4$ powders[25]. These targets were then used for high-pressure-oxygen sputter deposition of PYCCO films between 9 nm (~24 u.c.) and 150 nm (~390 u.c.) thickness, at similar conditions to prior work[33,41,43,49]. Briefly, substrates, such as YAO(101), were first annealed at 900 °C in 0.45 Torr of flowing ultrahigh-purity $O_2$ (99.998 %) for 15 mins prior to growth. Deposition then took place at 600 °C substrate temperature, ~30 W of DC sputter power, and 1.4 Torr of flowing $O_2$, yielding ~4.1 Å min$^{-1}$ growth rates; post growth, films were cooled to ambient in 600 Torr of $O_2$ at ~15 °C/min. Film thicknesses were determined by GIXR using a Rigaku SmartLab XE with Cu $K_\alpha$ ($\lambda$ = 1.5406 Å) radiation. Film microstructures and lattice parameters were probed by HRXRD, in specular and reciprocal space mapping modes, in the same system.

DC transport measurements were performed in a Quantum Design Physical Property Measurement System (PPMS), from 3 K to 300 K, using a Keithley 2400 source-measure unit. A four-terminal van der Pauw geometry was employed, with In contacts, carefully selecting excitation currents to avoid non-ohmicity and/or self-heating. Magnetometry was done in a Quantum Design Magnetic Property Measurement System and a Quantum Design PPMS with a vibrating sample magnetometer, from 5 K to 200 K, in in-plane magnetic fields to 70 kG (7 T); field cooling was performed in 10 kG (1 T). For magnetometry of films on SLAO, the substrate background was subtracted based on a three-sample-averaged measurement of nominally identical substrates at identical measurement conditions. Chemical and magnetic depth profiles were obtained from PNR at 5 K and 200 K in an in-plane 30 kG (3 T) field, on the Polarized Beam Reflectometer at the NIST Center for Neutron Research. $R^{++}$ and $R^{--}$ were measured vs. $Q_z$, where the "+" and "-" denote incident (and reflected) neutron spin polarization parallel or antiparallel to the magnetic field at the sample. PNR data were reduced using *Reductus*[60] software and fit to a slab model using *Refl1D*[61,62]. The fitting procedure included data at both temperatures, keeping all but the magnetic parameters constant. The uncertainty of the reflectivity is reported to one standard deviation, and model-fit parameter uncertainties to two standard deviations.

MFM was performed in an Attocube cantilever-based cryogenic atomic force microscope with interferometric detection, from 2 K to 295 K in applied magnetic fields up to 60 kG (6 T). Nanosensors PPP-MFMR probes were used, with a hard magnetic coating on the probe tip. The presented MFM images are the shift of the cantilever resonant frequency ($\Delta f$) and were measured with cantilever oscillation amplitudes from 40 nm to 85 nm. The tip-sample bias was chosen to minimize the electrostatic force, except where noted. Constant-height MFM images were taken with a fixed tip height, after correcting for the global tilt of the sample surface. Constant-lift MFM images were taken by first measuring the sample topography along one line, then repeating the line at a fixed offset to record $\Delta f$. In this way, topographic and MFM images were recorded in an interleaved fashion with negligible drift between the two. All out-of-plane applied magnetic field values shown for PYCCO films were high enough to align the tip magnetization with the field, so we do not distinguish between directions (+/-) of the applied field, as they are equivalent. Line-by-line linear subtraction or two-dimensional polynomial subtraction was used on the MFM and topographic images to remove background variations due to instrument drift and noise.

## DATA AVAILABILITY

Data supporting the results of this study are either provided in the paper or Supplementary Information, or are available from the corresponding author upon reasonable request.

**ACKNOWLEDGMENTS**

We thank P. Schiffer (Princeton University) for productive interactions regarding this work. Certain commercial products are identified here to describe our study adequately. Such identification is not intended to imply recommendation or endorsement by NIST.

**AUTHOR CONTRIBUTIONS**

JED and CL conceived of the study. JED, VC, WMP, LF, CK, and AJ deposited and structurally characterized the films, under the supervision of CL. Electronic and magnetic measurements and analyses were performed by JED, VC, and LF under the supervision of CL. PNR measurements were made by PQ, PPB, and BJK, and the data were analyzed by PQ, PPB, BJK and JED. MFM measurements and analyses were done by TAW and ANP. Theoretical analysis was done by PS, TB, and RMF. JED and CL led the overall interpretation of the results and wrote the paper with input from all authors.

**FUNDING**

Work at the University of Minnesota (UMN) was primarily supported by the Department of Energy (DOE) through the UMN Center for Quantum Materials under DE-SC0016371. Parts of this work were carried out in the Characterization Facility, UMN, which receives partial support from the National Science Foundation through the MRSEC program (under DMR-2011401). MFM measurements were supported by the Air Force Office of Scientific Research *via* grant FA9550-21-1-0378 (TAW) and by Programmable Quantum Materials, an Energy Frontier Research Center funded by the U.S. Department of Energy (DOE), Office of Science, Basic Energy Sciences (BES), under award DE-SC0019443 (ANP).

**COMPETING INTERESTS**

The authors declare no competing interests.

**ADDITIONAL INFORMATION**

**Supplementary information.** The online version contains supplementary material available at XXX.

**Correspondence** and requests for materials should be addressed to C. Leighton.

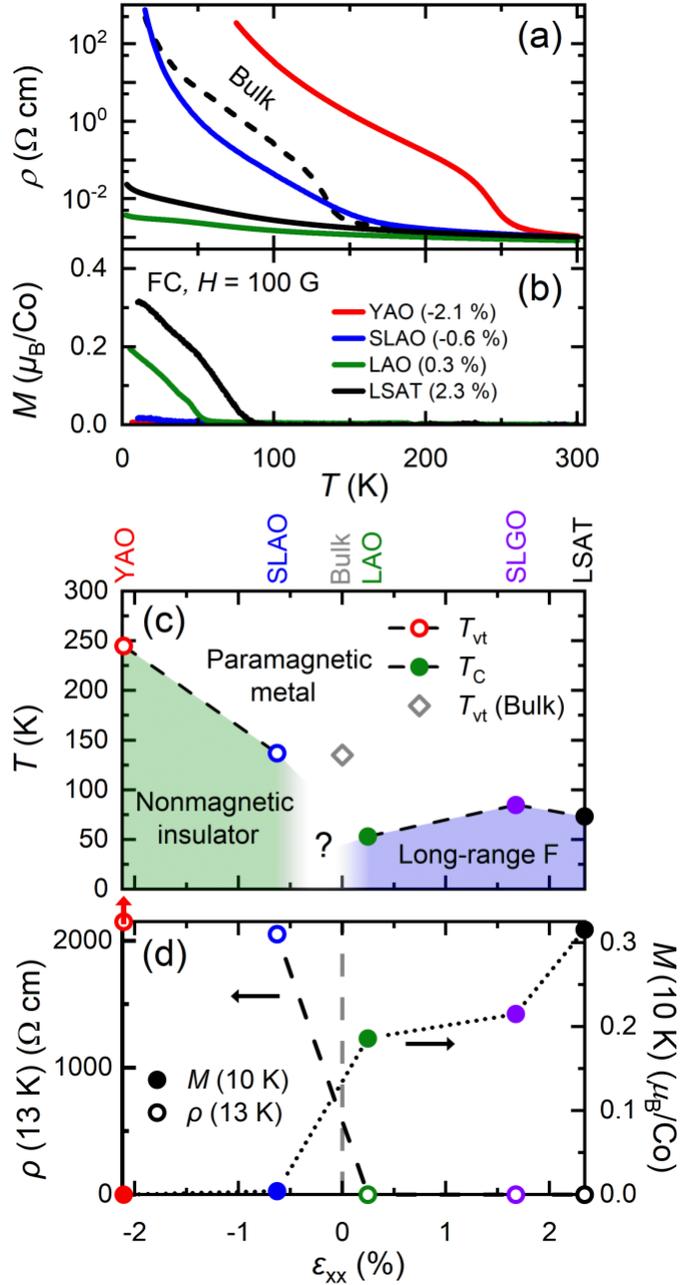

**Figure 1: Summary of the electronic and magnetic phase behavior of strain-tuned epitaxial (Pr$_{0.85}$Y$_{0.15}$)$_{0.7}$Ca$_{0.3}$CoO$_{3-\delta}$.** Temperature ($T$) dependence of (a) the resistivity ($\rho$, log$_{10}$ scale) and (b) the in-plane magnetization ($M$) of ~30 unit cell thick (Pr$_{0.85}$Y$_{0.15}$)$_{0.7}$Ca$_{0.3}$CoO$_{3-\delta}$ films on YAO(101), SLAO(001), LAO(001), SLGO(001), and LSAT(001). In (a), a bulk polycrystalline sample at the same composition is shown for reference (black dotted line). Both (a) and (b) were taken on warming, (b) in an in-plane applied field of 100 G (10 mT) after field cooling at 10 kG (1 T). (c) Experimental temperature ($T$) vs. "in-plane strain" ($\varepsilon_{xx}$) phase diagram for $(Pr_{0.85}Y_{0.15})_{0.7}Ca_{0.3}CoO_{3-\delta}$ films. Valence transition temperatures $T_{vt}$ (open circles) and Curie temperatures $T_C$ (filled circles) are plotted. Green, white, and blue phase fields indicate "nonmagnetic insulator", "paramagnetic metal", and "long-range ferromagnet (F)", respectively. Thin-film data are color coded by substrate (indicated at the top), and a polycrystalline bulk sample is shown (grey open diamond). (d) In-plane-strain dependence of the 13 K resistivity (left axis, open circles) and 10 K magnetization (right axis, closed circles, applied in-plane field of 100 G (10 mT)). Thin-film data are color coded by substrate in the same manner as (c). In (c) and (d), black dashed and dotted lines are guides to the eye. Panel (c) is adapted from Ref. [33] with permission.

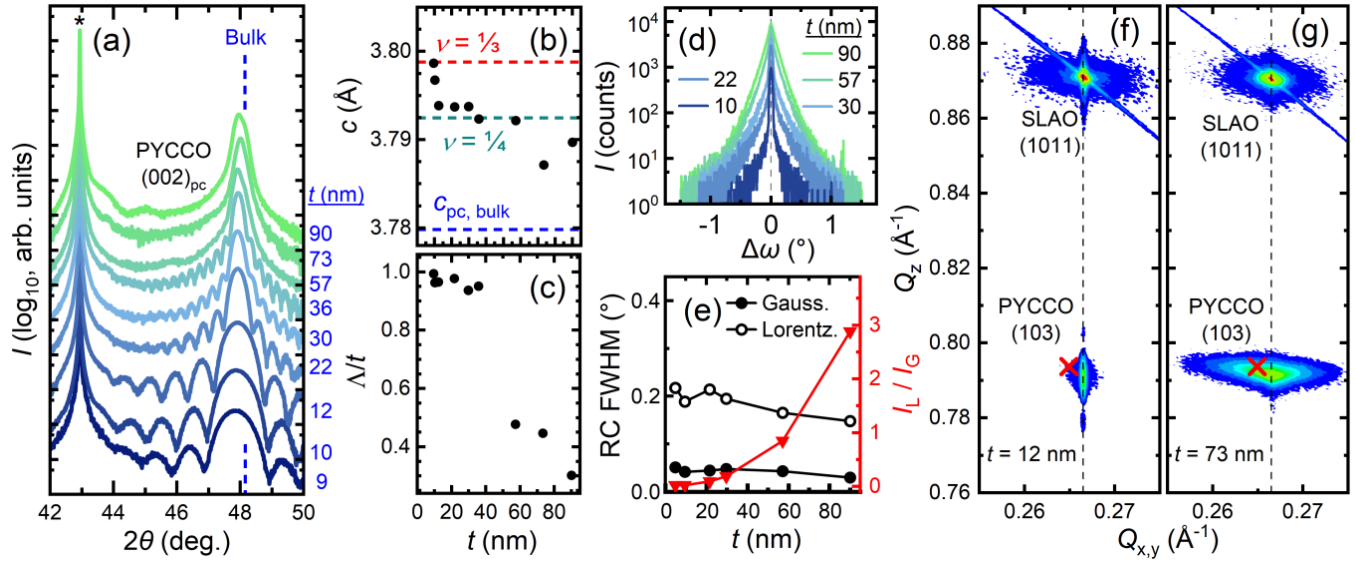

**Figure 2: Structural characterization of SrLaAlO$_4$(001)/(Pr$_{0.85}$Y$_{0.15}$)$_{0.7}$Ca$_{0.3}$CoO$_{3-\delta}$.** (a) Specular high-resolution X-ray diffraction around the pseudocubic 002 film peaks of 9 to 90 nm thick films. Scans are vertically offset for clarity, and labeled (to the right) with their thickness ($t$). Substrate 006 reflections are labeled "*", and the bulk pseudocubic 002 position is indicated (dashed blue line). Thickness dependence of (b) the out of plane lattice parameter ($c$), and (c) the Scherrer length normalized to the thickness ($\Lambda/t$). In (b), the expected fully-strained value (red and green dashed lines, based on a Poisson ratio ($\nu$) between 1/3 and 1/4) and bulk pseudocubic value (blue dashed line) are indicated. (d) X-ray rocking curves (intensity $I$ vs. rocked incident angle $\Delta\omega$) taken about the pseudocubic 002 film peaks at $t$ = (10, 22, 30, 57, and 90) nm. (e) $t$ dependence of the rocking curve (RC) full-width-half-maxima (FWHM) (left axis), fit as the sum of Gaussian (closed circles) and Lorentzian (open circles) peaks. The ratio of Lorentzian and Gaussian intensities ($I_L/I_G$, red triangles) is indicated on the right axis. (f,g) Asymmetric X-ray reciprocal space maps around the pseudocubic 103 film peaks of 12 nm thick and 73 nm thick films. Bulk (relaxed) positions are marked with a red ×, while vertical black dashed lines correspond to full strain.

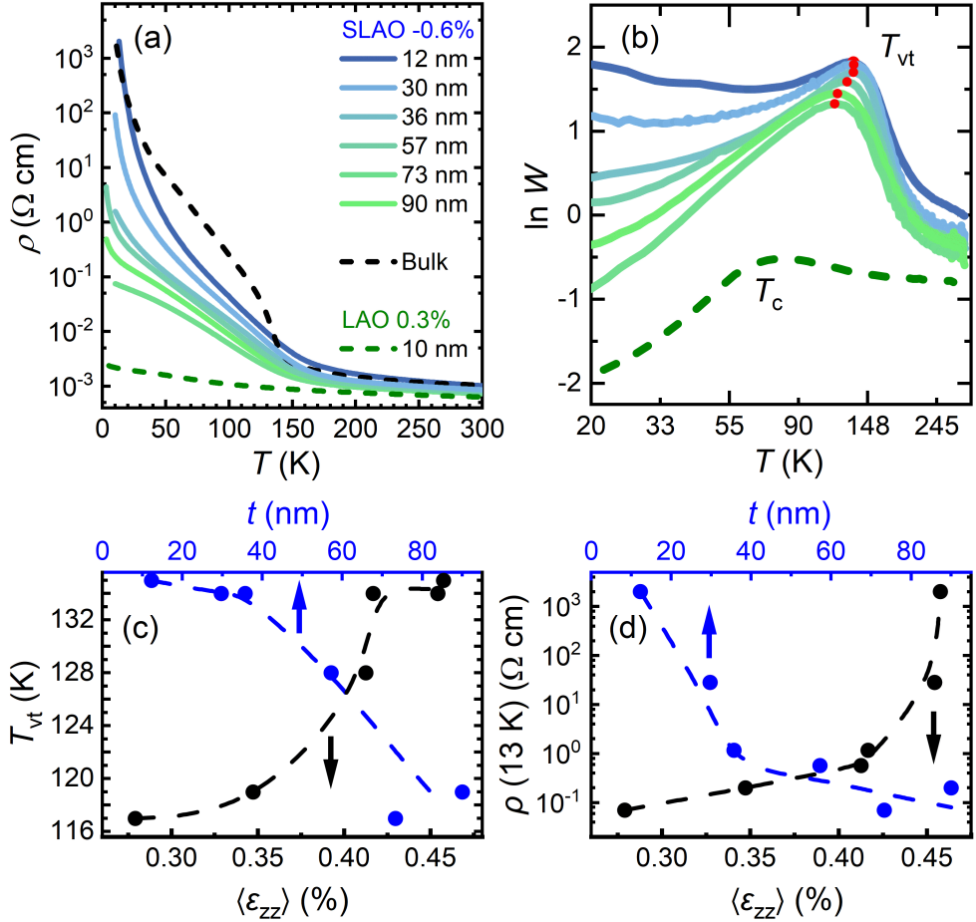

**Figure 3: Thickness-dependent transport properties of SrLaAlO$_4$(001)/ (Pr$_{0.85}$Y$_{0.15}$)$_{0.7}$Ca$_{0.3}$CoO$_{3-\delta}$.** Temperature ($T$) dependence of the resistivity ($\rho$) (a, log$_{10}$ scale, taken on warming) and corresponding Zabrodskii plots[50] (b, ln $W$ vs. ln $T$, where $W = -d\ln\rho/d\ln T$) of 12 nm thick to 90 nm thick films. For reference, a fully strained film on LAO (+0.3 % in-plane strain, 10 nm thickness) (dashed green lines) and a bulk polycrystalline sample (black dashed line) are included. Valence transition temperatures ($T_{vt}$) are highlighted in (b) by red dots. (c,d) Thickness ($t$, top axes) and "out-of-plane" strain ($\langle\varepsilon_{zz}\rangle$, bottom axes) dependence of $T_{vt}$ (c) and $\rho$ at $T$ = 13 K (d), as extracted from the data shown in (a,b); dashed lines are guides to the eye.

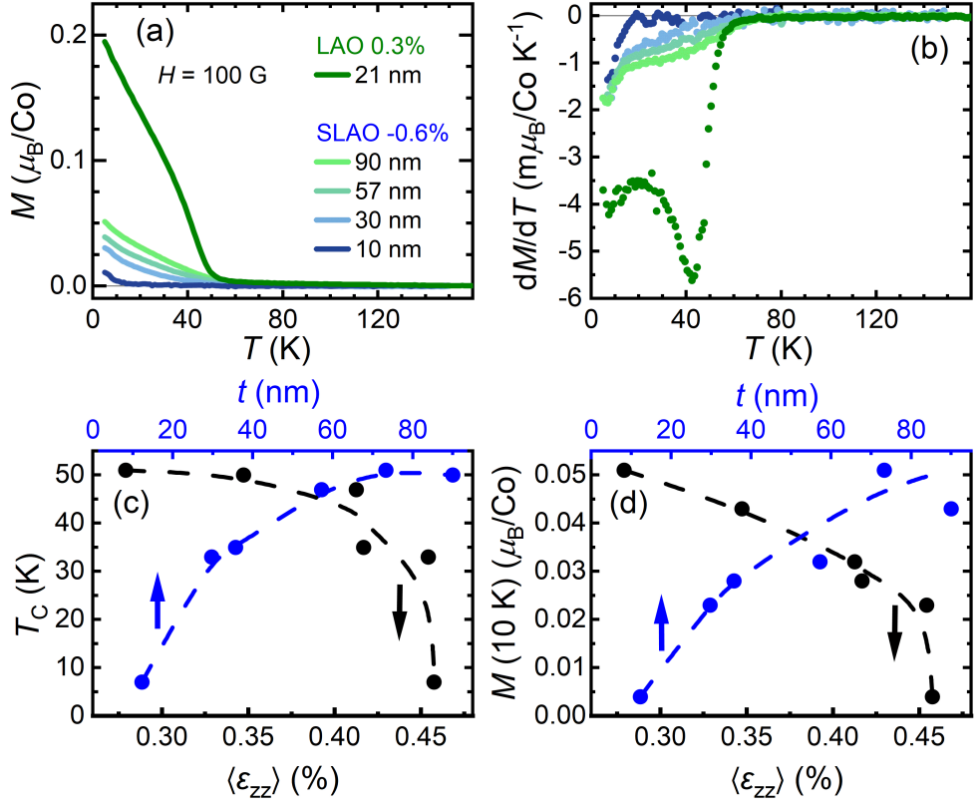

**Figure 4: Thickness-dependent magnetic properties of SrLaAlO$_4$(001)/ (Pr$_{0.85}$Y$_{0.15}$)$_{0.7}$Ca$_{0.3}$CoO$_{3-\delta}$.** Temperature ($T$) dependence of the magnetization ($M$) (a) and corresponding temperature derivative ($dM/dT$) (b) of 10 nm thick to 90 nm thick films. For reference, a fully strained film on LAO (+0.3 % in-plane strain, 21 nm thickness) (dark green line and circles) is included. (c,d) Thickness ($t$, top axes) and "out-of-plane" strain ($\langle\varepsilon_{zz}\rangle$, bottom axes) dependence of $T_C$ (c) and $M$ at $T = 10$ K (d), as extracted from the data shown in (a); dashed lines are guides to the eye. All magnetometry data are in 100 G (10 mT) after field cooling in 10 kG (1 T).

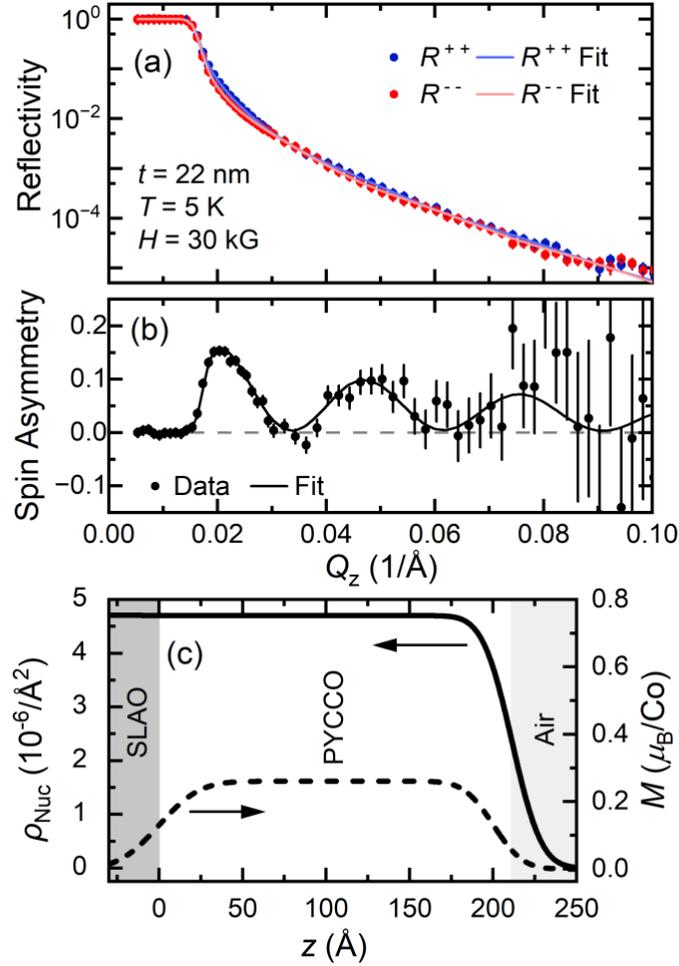

**Figure 5: Polarized neutron reflectometry (PNR) characterization of SrLaAlO$_4$(001)/ (Pr$_{0.85}$Y$_{0.15}$)$_{0.7}$Ca$_{0.3}$CoO$_{3-\delta}$.** (a) Polarized neutron reflectivity *vs.* scattering wave vector magnitude ($Q_z$) from a 22 nm thick film at 5 K in a 30 kG (3 T) in-plane magnetic field ($H$). Blue and red points denote the non-spin-flip channels $R^{++}$ and $R^{--}$, respectively, and the solid lines are the fits discussed in the text. (b) Spin asymmetry [SA = ($R^{++}$-$R^{--}$)/($R^{++}$+$R^{--}$)] *vs.* $Q_z$ extracted from (a). (c) Depth ($z$) profiles of the nuclear scattering length density ($\rho_{nuc}$, left-axis) and magnetization ($M$, right-axis) extracted from the fits to the data shown in (a,b). See Supplementary Table 1 for all reflectometry fit parameters. Error bars in (a) and (b) are ±1 standard deviation.

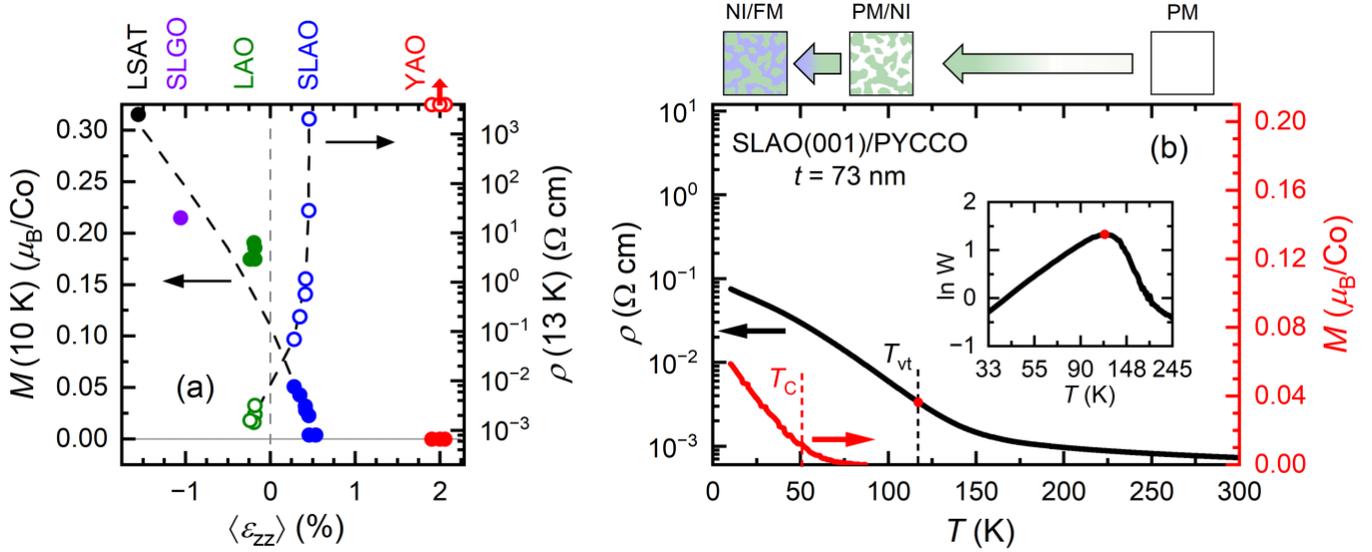

**Figure 6: Strain-dependent electronic/magnetic phase transition and phase coexistence in (Pr$_{0.85}$Y$_{0.15}$)$_{0.7}$Ca$_{0.3}$CoO$_{3-\delta}$ films.** (a) "Out-of-plane" strain ($\langle \varepsilon_{zz} \rangle$) dependence of 10 K field-cooled magnetization ($M$, left-axis, closed circles, in an in-plane field of 100 G (10 mT)) and 13 K resistivity ($\rho$, right axis, log$_{10}$ scale, open circles) for fully-strained and partially-strain-relaxed (Pr$_{0.85}$Y$_{0.15}$)$_{0.7}$Ca$_{0.3}$CoO$_{3-\delta}$ on YAO(101), SLAO(001), LAO(001), SLGO(001), and LSAT(001). Data are color coded by substrate (indicated at the top), and dashed lines are guides to the eye. (b) Temperature ($T$) dependence of resistivity (left axis, log$_{10}$ scale), and magnetization (right axis, in an in-plane field of 100 G (10 mT)) of a 73 nm thick film on SLAO(001). Valence transition and Curie temperatures ($T_{vt}$ and $T_C$, respectively) are indicated by dashed lines, and $T_{vt}$ is highlighted by a red dot. Inset: corresponding Zabrodskii plot[51] (ln $W$ vs. ln $T$, where $W = -d\ln\rho/d\ln T$). The hypothesized spatial arrangement of electronic/magnetic phases is schematically illustrated above (b), from "paramagnetic metal" (PM) at room-$T$ to mixed-phase PM and "nonmagnetic insulator" (NI) below $T_{vt}$, to mixed-phase NI and "ferromagnetic metal" (FM) below $T_C$.

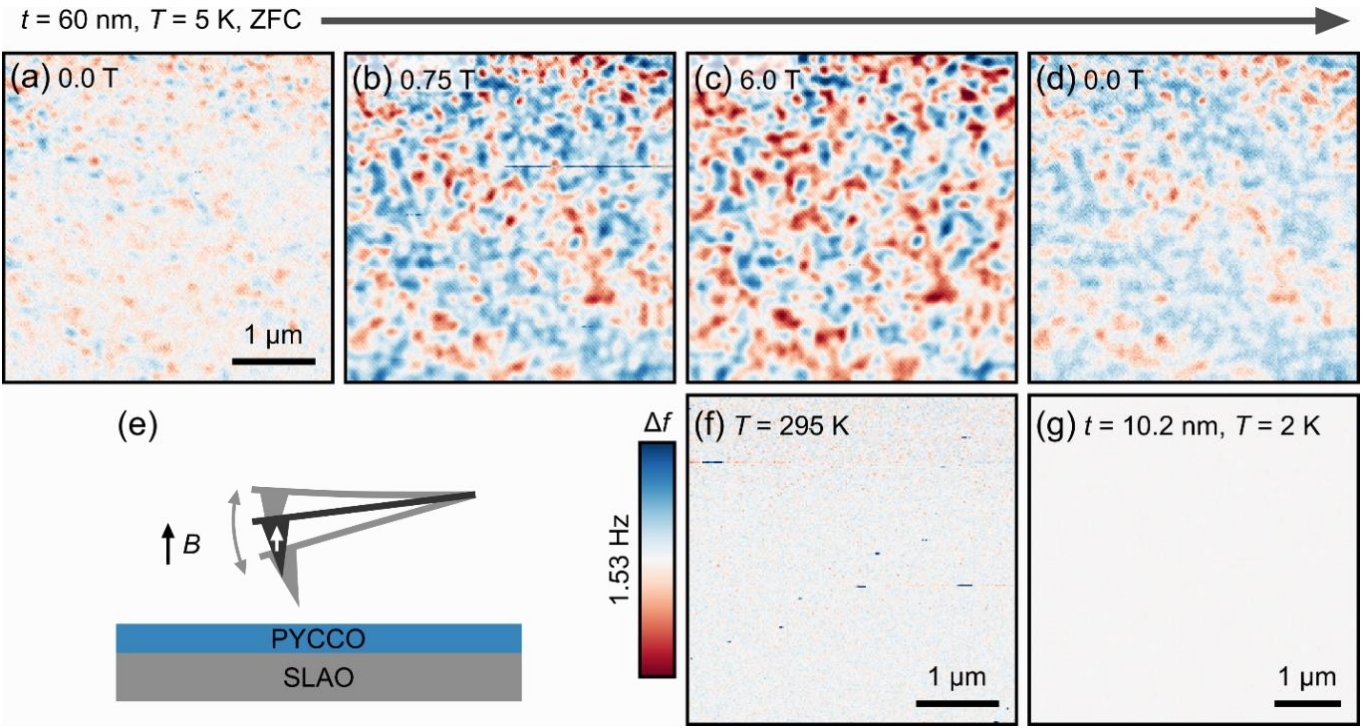

**Figure 7: Temperature-dependent magnetic force microscopy (MFM) measurements on SrLaAlO$_4$(001)/(Pr$_{0.85}$Y$_{0.15}$)$_{0.7}$Ca$_{0.3}$CoO$_{3-\delta}$ films.** (a-d) 5 K MFM images of a partially-strain-relaxed SLAO(001)/(Pr$_{0.85}$Y$_{0.15}$)$_{0.7}$Ca$_{0.3}$CoO$_{3-\delta}$ (60 nm) film after zero-field cooling (ZFC). The out-of-plane applied magnetic field increases with the arrow and is: (a) 0, (b) 7.5 kG (0.75 T), (c) 60 kG (6 T), and (d) returned to 0. (e) MFM measurement schematic. The MFM images here were collected at a constant height of ~50 nm above the film surface, with the applied field and tip magnetization oriented out-of-plane and a tip-sample bias of 0 V. All images are on the same color scale (see left side of panel (f)) for $\Delta f$, the shift of the cantilever resonant frequency. (f) MFM image of the same film as in (a-d) using the same probe tip, at zero field and 295 K (*i.e.*, far above the Curie temperature of ~45 K (Fig. 4)); no magnetic contrast is distinguishable. (g) MFM image of a SLAO(001)/(Pr$_{0.85}$Y$_{0.15}$)$_{0.7}$Ca$_{0.3}$CoO$_{3-\delta}$(10.2 nm) film at zero field and 2 K (after ZFC), for comparison; no magnetic contrast is distinguishable. This image was taken with a different probe tip from (a-d) but with comparable properties.

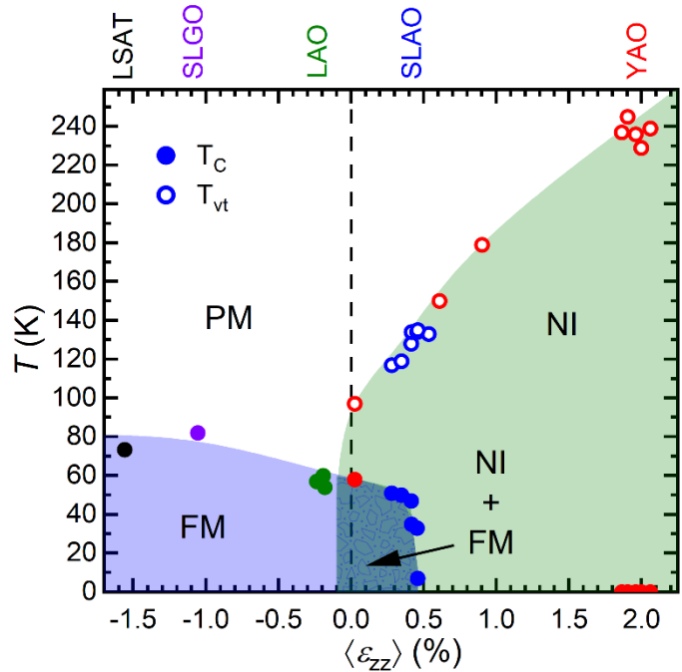

**Figure 8: Final "strain phase diagram" of $(Pr_{0.85}Y_{0.15})_{0.7}Ca_{0.3}CoO_{3-\delta}$.** Temperature ($T$) *vs*. "out-of-plane strain" ($\varepsilon_{zz}$) phase diagram for pseudomorphic and partially-strain-relaxed $(Pr_{0.85}Y_{0.15})_{0.7}Ca_{0.3}CoO_{3-\delta}$ films. Valence transition temperatures $T_{vt}$ (open circles) and Curie temperatures $T_C$ (filled circles) are plotted. Green, white, and blue phase fields indicate "nonmagnetic insulator" (NI), "paramagnetic metal" (PM), and "ferromagnetic metal" (FM), respectively. Data are color coded by substrate (indicated at the top).

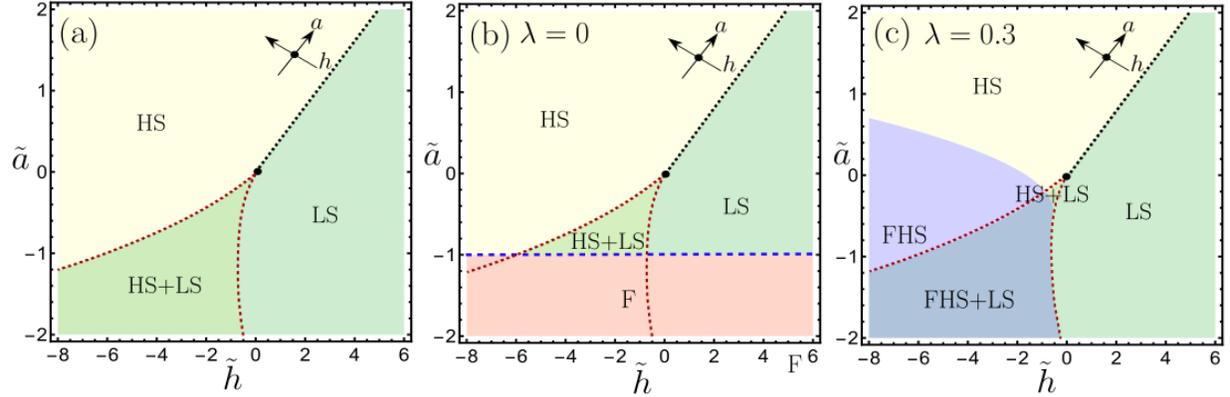

**Figure 9: Phase diagrams of the phenomenological Landau model for the coupled spin-state and ferromagnetic transitions**. (a) Refers only to the spin-state phase diagram, (b) to the uncoupled spin-state and magnetic phase diagrams (*i.e.*, $\lambda = 0$), and (c) to the coupled spin-state and magnetic phase diagram with $\lambda = 0.3$. The Landau parameters $\tilde{a}$ and $\tilde{h}$ depend only on temperature and strain *via* $\tilde{a} = \tilde{a}_0(T - T_{st})$ and $\tilde{h} = \tilde{h}_1(\varepsilon - \varepsilon_{st})$, such that the spin-state critical endpoint (black circle) occurs at $\tilde{a} = \tilde{h} = 0$. The black dotted line indicates a crossover between the low-spin (LS, green) and high-spin (HS, yellow) phases, whereas the red dotted lines are the spinodal lines of metastability of the two phases. In (b), the horizontal dashed blue line corresponds to the strain-independent bare Curie temperature, giving rise to ferromagnetic order (F, orange) that is uncoupled from the spin state. In (c), FHS (purple) denotes the ferromagnetic high-spin state. The parameters used for all of these plots are $\alpha_1 = 4, \beta_1 = 0.5, \alpha_2 = 5, \beta_2 = -2, \Delta t = 1$.

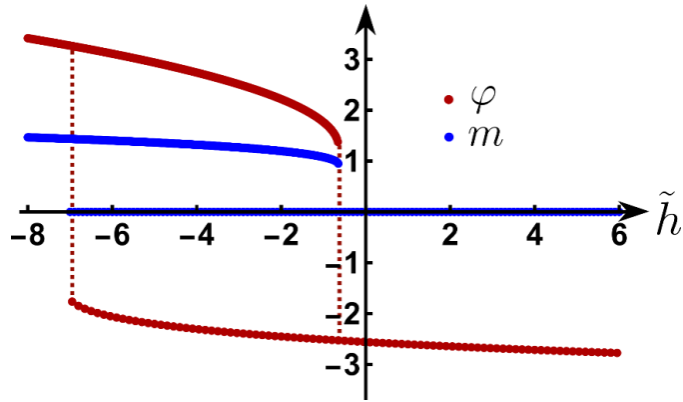

**Figure 10: Order parameters as a function of $\tilde{h}$ for fixed $\tilde{a}$.** Shown are the magnetic order parameter $m$ (blue) and spin-state order parameter $\varphi$ (red) as a function of $\tilde{h}$ for $\tilde{a} = -1.1$ and $\lambda = 0.3$. Note the region of coexistence where both positive and negative $\varphi$ appear as local minima of the free energy.

Supplementary Information for

# First-Order Phase Transition *vs*. Spin-State Quantum-Critical Scenarios in Strain-Tuned Epitaxial Cobaltite Thin Films

John E. Dewey[1], Vipul Chaturvedi[1], Tatiana A. Webb[2], Prachi Sharma[3], William M. Postiglione[1], Patrick Quarterman[4], Purnima P. Balakrishnan[4], Brian J. Kirby[4], Lucca Figari[1], Caroline Korostynski[1], Andrew Jacobson[1], Turan Birol[1], Rafael M. Fernandes[3], Abhay N. Pasupathy[2,5] and Chris Leighton[1]*

[1]Department of Chemical Engineering and Materials Science, University of Minnesota, Minneapolis, Minnesota 55455, USA

[2]Department of Physics, Columbia University, New York, New York 10027, USA

[3]School of Physics and Astronomy, University of Minnesota, Minneapolis, Minnesota 55455, USA

[4]NIST Center for Neutron Research, National Institute of Standards and Technology, Gaithersburg, Maryland 60439, USA

[5]Condensed Matter Physics and Materials Science Division, Brookhaven National Laboratory, Upton, New York 11973, USA

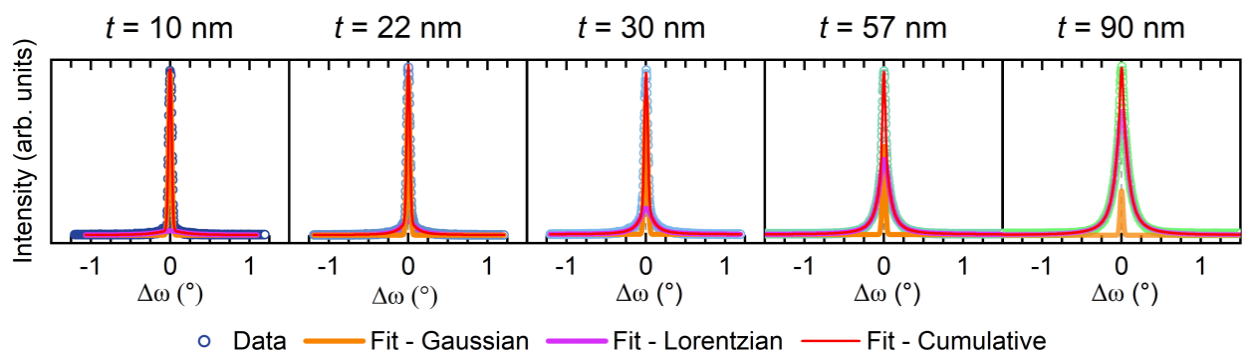

**Figure S1: Peak fitting of SrLaAlO$_4$(001)/(Pr$_{0.85}$Y$_{0.15}$)$_{0.7}$Ca$_{0.3}$CoO$_{3-\delta}$ X-ray rocking curves.** X-ray rocking curves (intensity $I$ *vs.* rocked incident angle $\Delta\omega$) through pseudocubic 002 film peaks for films with thickness ($t$) of 10 nm, 22 nm, 30 nm, 57 nm, and 90 nm, as in the main text. For the data, the color scheme is the same as Fig. 2 of the main text. The solid red lines are fits composed of the sum of a Gaussian peak (orange solid lines) and a Lorentzian peak (purple solid lines).

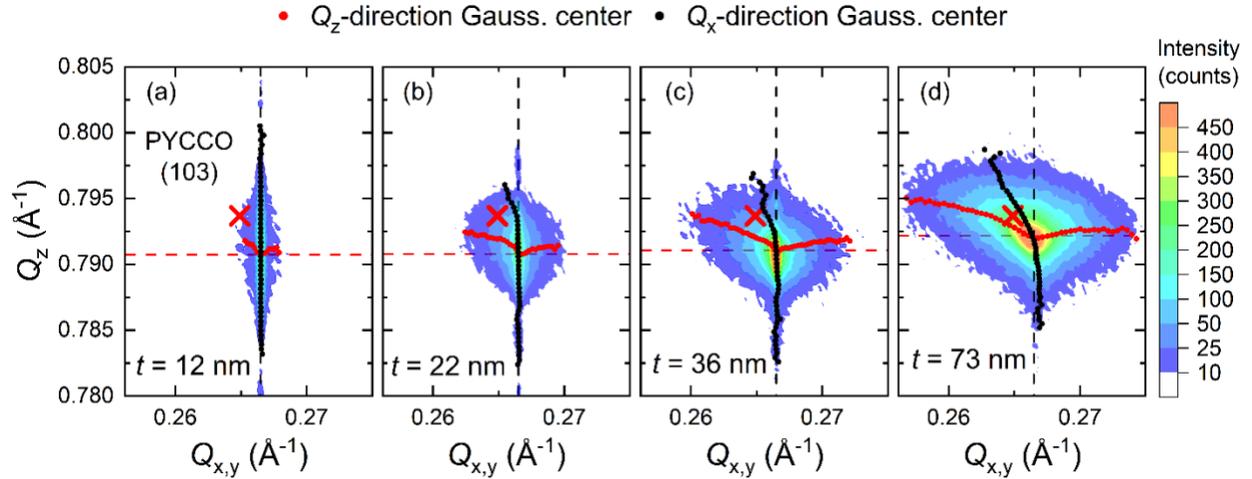

**Figure S2: Additional X-ray reciprocal space mapping of SrLaAlO$_4$(001)/ (Pr$_{0.85}$Y$_{0.15}$)$_{0.7}$Ca$_{0.3}$CoO$_{3-\delta}$ films**. Asymmetric X-ray reciprocal space maps around the pseudocubic (pc) 103 film peaks of films with thickness of 12 nm (a), 22 nm (b) 36 nm (c), and 73 nm (d). Bulk (relaxed) positions are marked with a red ×, while vertical black dashed lines correspond to full strain, *i.e.*, pseudomorphic growth. Horizontal red dashed lines correspond to the out-of-plane lattice parameter measured *via* high resolution specular X-ray diffraction at the pseudocubic 002 film peak (Fig. 2(a)), converted to $3/d_{001}$ ($d$ = interplanar spacing). The intensity (*I*) peaks in (a-d) were fit to an orthogonal grid of 1D Gaussians, along either the out-of-plane scattering vector ($Q_z$) or the in-plane vector ($Q_{x,y}$). The Gaussian peak centers of each fit are labeled with red dots (for $I(Q_z)$ fits at constant $Q_{x,y}$) or black dots (for $I(Q_{x,y})$ fits at constant $Q_z$), generating the solid red and black lines shown.

**Note:** As described in the main text, an attempt was made to simultaneously extract in-plane (IP) and out-of-plane (OoP) lattice parameters ($a_{pc}$, $c_{pc}$) from the data in Fig. S2 to generate a "master curve" of PYCCO $a_{pc}(t)$ and $V_{uc}(t)$ (the latter being $a_{pc}(t)^2 c_{pc}(t)$). The conventional expectation is that, upon partial relaxation of compressive IP strain, the OoP lattice parameter should contract, and the 103$_{pc}$ peak positions in Figs. 2(g) and S2(b-d) should move away from the pseudomorphic crystal truncation rod (vertical black dashed line), towards the bulk PYCCO position (red ×). To determine $a_{pc}(t)$ and $c_{pc}(t)$, the 103$_{pc}$ peak maxima in Figs. S2(a-d) were first located *via* orthogonal 1D Gaussian fitting, *i.e.*, locating fitted $I(Q_{x,y})$ peak centers at fixed $Q_z$ (black points in Fig. S2), followed by locating fitted $I(Q_z)$ peak centers at fixed $Q_{x,y}$ (red points in Fig. S2). The intersection of black and red points in each panel indicates the position of maximum film peak intensity. The

OoP lattice parameter extracted from these RSM peak positions ($c = 3d_{003}$) is in perfect agreement with the OoP lattice parameters measured by specular WAXRD in Fig. 2 ($c = 2d_{002}$), as demonstrated in Fig. S2 by the intersection of black points, red points, and horizontal red dashed lines (located at $Q_z = 3/2d_{002} = 1/d_{003}$). These intersection points - the peak "centers" - shift toward higher $Q_z$ at higher $t$ (Figs. S2(c,d)), indicating progression to a smaller OoP lattice parameter, as expected. Surprisingly, however, the peak "centers" at high $t$ in Fig. S2 do *not* shift in $Q_{x,y}$. Instead of moving toward the relaxed IP position (same $Q_{x,y}$ as the red × in Figs. S2(a-d)), the IP film peak coordinate remains pinned to the pseudomorphic position (vertical black dashed line), even at $t =$ 73 nm (Fig. S2(d)). Naïve application of Bragg's law to the peak maxima $Q_{x,y}$ and $Q_z$ in Figs. 2(f,g) and S2 would imply a decreasing out-of-plane lattice parameter with $t$ (higher peak $Q_z$), without an increase of the in-plane lattice parameter (lower peak $Q_{x,y}$), *i.e.*, an *intensification* of volumetric strain at high $t$, which would be unphysical. Even though the RSM peak *maximum* position does not deviate from the pseudomorphic $Q_{x,y}$ at high $t$, it is reasonable to assume that the (average) film in-plane lattice parameter must increase at high $t$. This is qualitatively indicated in Fig. S2 by the monotonic increase in off-pseudomorphic intensity with increasing $t$, streaked toward the bulk position (red ×). The "pinning" of $103_{pc}$ peak $Q_{x,y}(t)$ could be the consequence of diffuse scattering near dislocations, which effectively reduces the diffracted $I$ from partially-strain-relaxed portions of the film (at lower $Q_{x,y}$) relative to coherently-strained portions (at pseudomorphic $Q_{x,y}$), as previously analyzed for XRD in the low-dislocation-density limit[1] and observed, *e.g.*, in XRD of partially-strain-relaxed films of MgO/CoO[2]. Faced with this resulting uncertainty in the determination of the average IP lattice parameter *vs*. $t$ (and thus $V_{uc}$ *vs*. $t$), we thus quantify the degree of strain relaxation in SLAO/PYCCO using the average OoP strain ($\langle\varepsilon_{zz}\rangle$), defined as $\langle\varepsilon_{zz}\rangle = [\langle c \rangle - c_{pc}]/c_{pc}$, where $\langle c \rangle$ is the $t$-dependent average $c$-axis lattice parameter measured by HRXRD in Fig. 2(b).

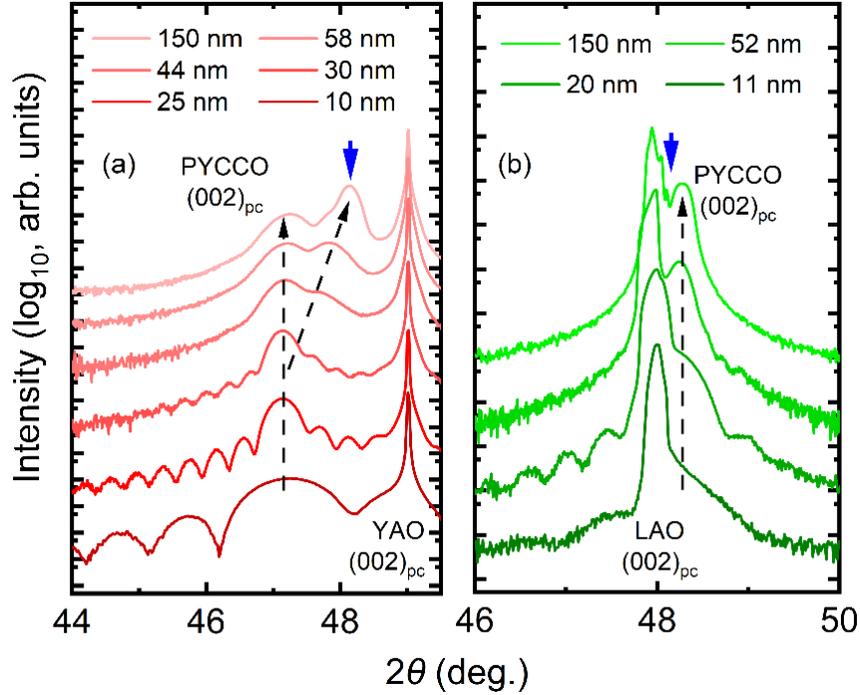

**Figure S3: Structural characterization of $(Pr_{0.85}Y_{0.15})_{0.7}Ca_{0.3}CoO_{3-\delta}$ films on $YAlO_3(101)$ and $LaAlO_3(001)$ substrates.** Specular high-resolution X-ray diffraction scans around the pseudocubic (pc) 002 film peaks of 10 nm to 150 nm films on $YAlO_3(101)$ (a), and 11 nm to 150 nm films on $LaAlO_3(001)$ (b). The scans are vertically offset for clarity. The bulk $002_{pc}$ position is indicated by blue arrows. Black arrows are guides to the eye, indicating the evolution of the $002_{pc}$ peak position with increasing thickness. In (a), the discontinuous split into two distinct peaks above the critical thickness is clear, as discussed in the main text. In (b), there is no detectable strain relaxation, as also discussed in the main text.

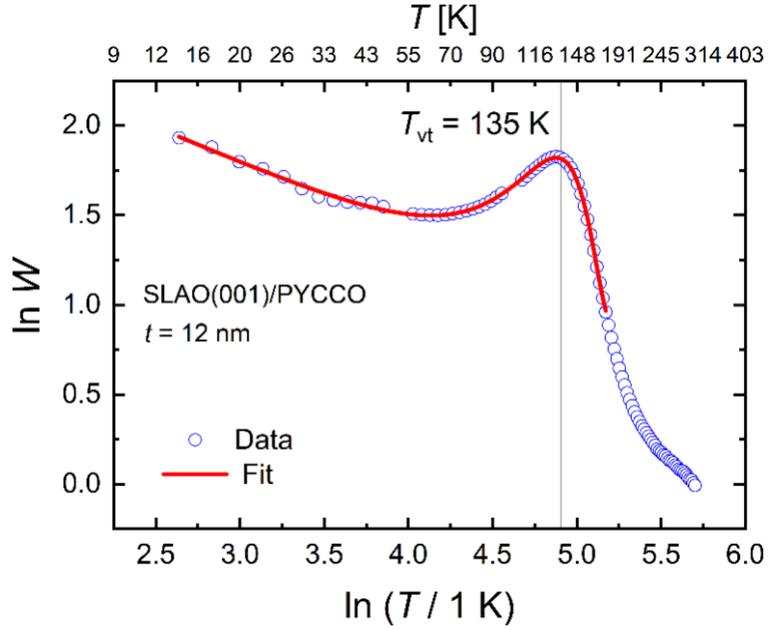

**Figure S4: Example fitting of SrLaAlO$_4$(001)/(Pr$_{0.85}$Y$_{0.15}$)$_{0.7}$Ca$_{0.3}$CoO$_{3-\delta}$ resistivity versus temperature in Zabrodskii format.** A representative fit of the temperature-dependent resistivity ($\rho(T)$, 12 nm thickness) in Zabrodskii format[3], ln $W$ vs. ln $T$, where $W = -d\ln\rho/d\ln T$. As discussed in the main text, a skewed asymmetric Gaussian (red line) is used to fit the data (blue circles), with the peak center denoting the valence transition temperature ($T_{vt}$), indicated by a vertical gray line.

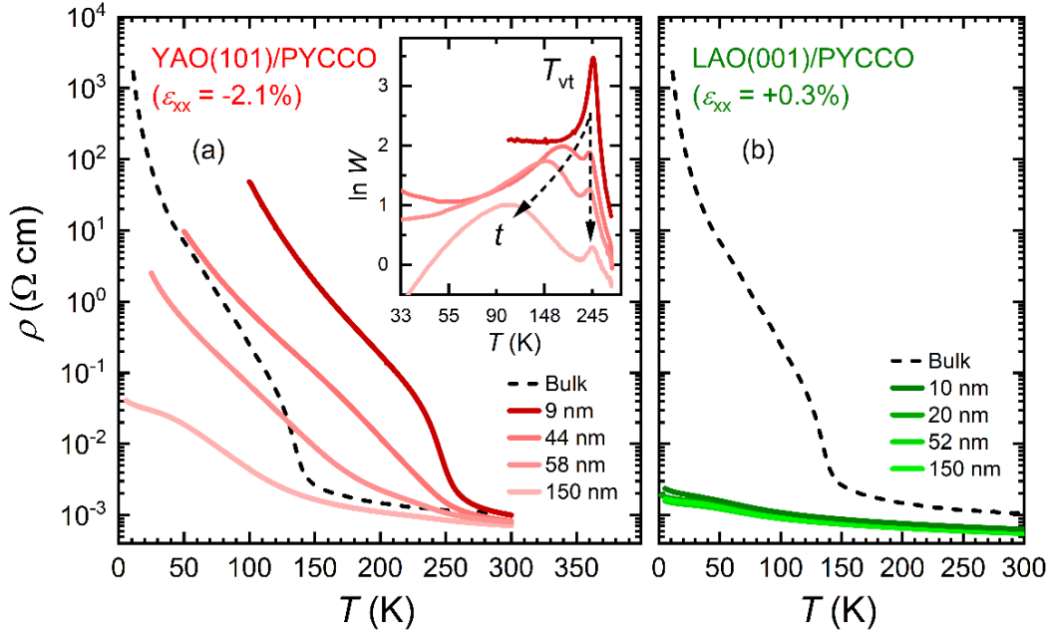

**Figure S5: Thickness-dependent electronic transport properties of $(Pr_{0.85}Y_{0.15})_{0.7}Ca_{0.3}CoO_{3-\delta}$ films on $YAlO_3(101)$ and $LaAlO_3(001)$ substrates.** Temperature ($T$) dependence of the resistivity ($\rho$) ($\log_{10}$ scale, taken on warming) of 9 nm to 150 nm films on $YAlO_3(101)$[4] (a), and 10 nm to 150 nm films on $LaAlO_3(001)$ (b). For reference, a bulk polycrystalline sample (black dashed line) is included. The inset in (a) is $\rho(T)$ data plotted in Zabrodskii format[3], $\ln W$ vs. $\ln T$, where $W = -d\ln\rho/d\ln T$. Peaks in the Zabrodskii plot indicate the valence transition temperature ($T_{vt}$), and black arrows indicate the evolution of $T_{vt}$ with increasing thickness ($t$), including the splitting into two distinct transitions, as discussed in the main text. In (b), consistent with Fig. S3(b), no thickness dependence is evident.

**Table S1:** Summary of parameters extracted from refinement of 5 K polarized neutron reflectometry data on 22 nm $(Pr_{0.85}Y_{0.15})_{0.7}Ca_{0.3}CoO_{3-\delta}$ (PYCCO) on SLAO (as shown in Fig. 5). The parameters are (from top to bottom) the nuclear scattering length density ($\rho_{Nuc}$), thickness ($t$), roughness ($\sigma$) (*i.e.*, Gaussian interface width), and magnetization ($M$) for each layer. Best fits were obtained with a magnetically-dead top (*i.e.*, surface) layer. This layer has zero magnetization but identical $\rho_{Nuc}$ to the bulk (middle) layer. Model parameter uncertainties represent ±2 standard deviations. As noted in the main text, the $\rho_{Nuc}$ values of SLAO and PYCCO are near-identical, which is the origin of the large uncertainty in $\sigma$ at the SLAO/PYCCO interface. As stated in Methods, similar measurements were also made at 200 K, where a small magnetization of 0.05 $\mu_B$/Co was found to persist (in 30 kG (3 T)).

| Sample | Substrate | Middle PYCCO | Top PYCCO |
|---|---|---|---|
| SLAO/PYCCO | $\rho_{Nuc} = 4.71 \times 10^{-6}$ Å$^{-2}$ | $\rho_{Nuc} = 4.70 \times 10^{-6}$ Å$^{-2}$ | $\rho_{Nuc} = 4.70 \times 10^{-6}$ Å$^{-2}$ |
| | $t = \infty$ | $t = 211.4$ Å ± 4.25 Å | $t = 10.1$ Å ± 8.1 Å |
| | $\sigma = 19.0$ Å ± 7.4 Å | $\sigma = 13.87$ Å ± 0.16 Å | - |
| | $M = 0$ | $M = (0.26 \pm 0.01)$ $\mu_B$/Co | $M = 0$ |

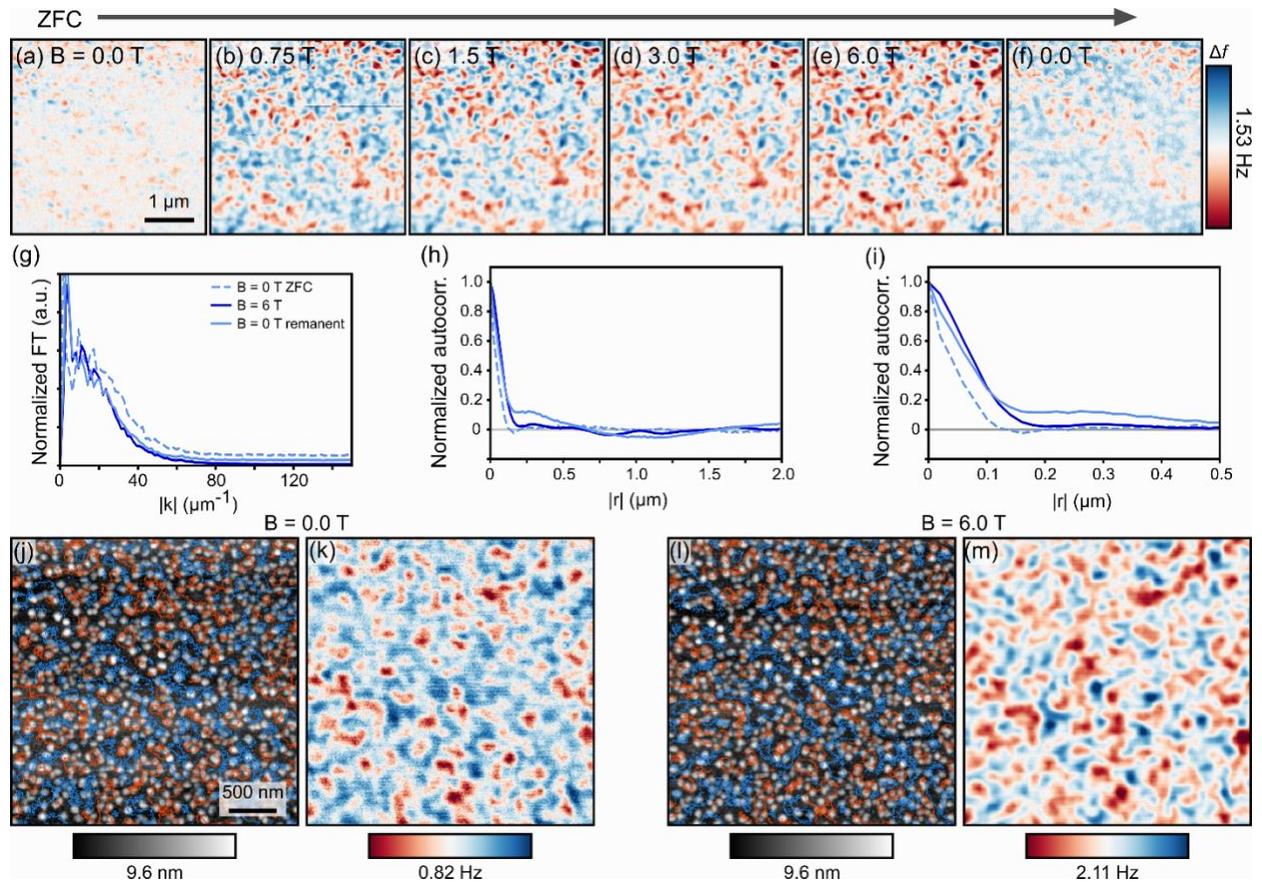

**Figure S6: Further characterization of magnetic inhomogeneity in the SrLaAlO$_4$(001)/ (Pr$_{0.85}$Y$_{0.15}$)$_{0.7}$Ca$_{0.3}$CoO$_{3-\delta}$(60 nm) film by magnetic force microscopy (MFM).** This is the same film as in Figs. 7(a-d,f) of the main text. (a-f) 5 K constant-height MFM images over the same nominal field of view, starting from a zero-field-cooled (ZFC) state. The out-of-plane applied magnetic field is (a) 0, (b) 7.5 kG (0.75 T), (c) 15 kG (1.5 T), (d) 30 kG (3 T), (e) 60 kG (6 T), and (f) returned to 0. Panels (a,b,e,f) are thus repeated from the main text. The tip height was 50 nm and the tip and sample biases were both 0 V. (a-f) are on the same color scale (right side of panel (f)) for $\Delta f$, the shift of the cantilever resonant frequency. (g) Angular-averaged amplitude of the Fourier transform (FT) of the images in (a,e,f), *i.e.*, at 0 T, 6 T, and 0 T, respectively. The broad peak indicates that the magnetic inhomogeneity in (a,e,f) exists across a broad range of wavelengths ($2\pi/k$). (h,i) Angular-averaged autocorrelation of the images in (a,e,f) over 2.0 μm (h), and 0.5 μm (i); the autocorrelations were normalized to 1 at $r = 0$. The correlation drops by half over 37 nm (a), 73 nm (e), and 64 nm (f). (j-m) 5-K constant-lift MFM (right panels) and simultaneously acquired topographic images (left panels), at zero field after ramping down from 6 T (j,k), and at 6 T (l,m). The 15$^{th}$- and 85$^{th}$-percentile contours of the MFM images are superimposed on the topographic images to demonstrate that while some of the MFM features may be correlated with topography, the overall features in the MFM images are not purely due to topography. The tip lift here is 30 nm and the tip and sample biases were 0 V. (k) and (m) each have their own color scale for $\Delta f$. (j) and (l) have the same height scale.

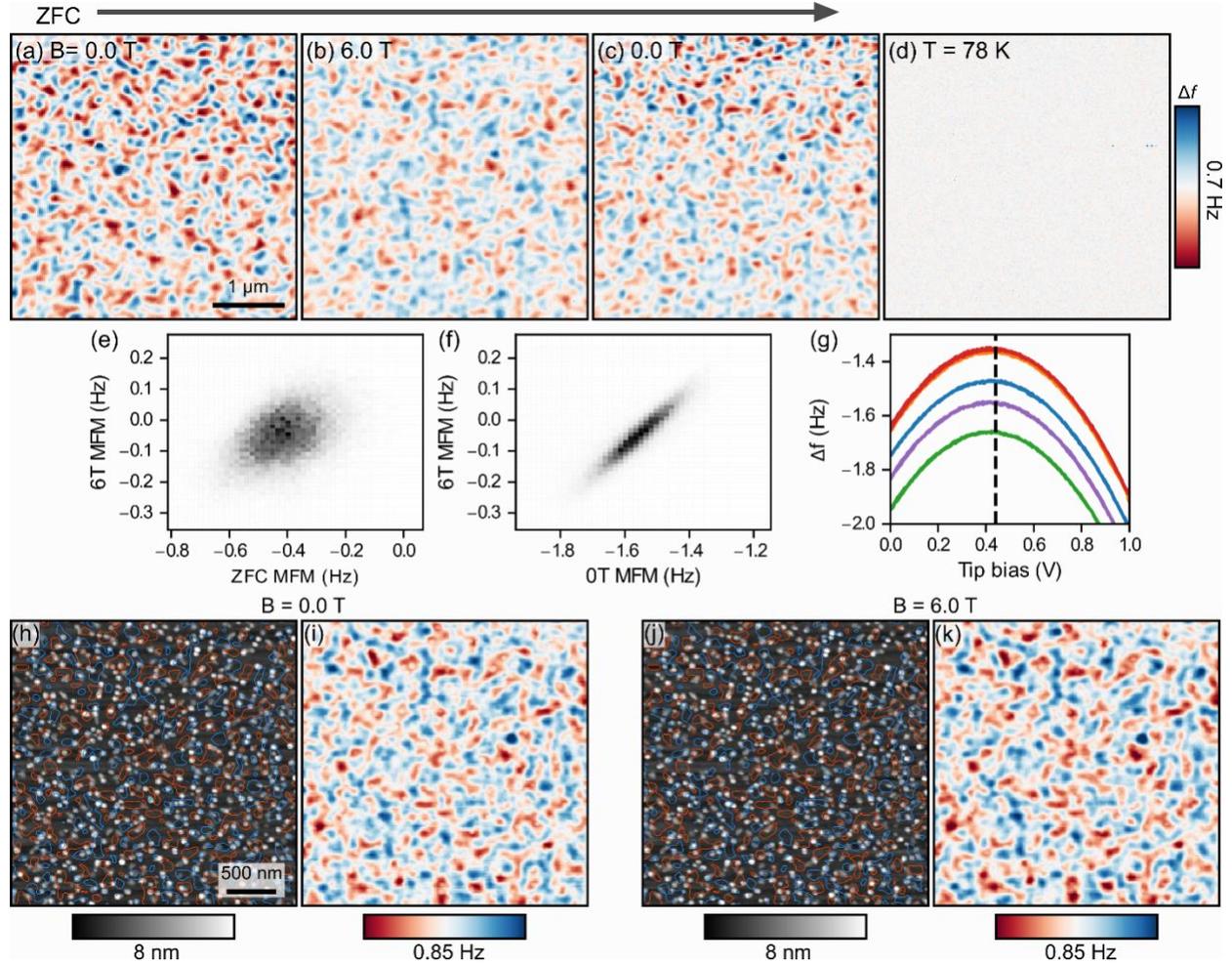

**Figure S7: Comparative characterization of magnetic inhomogeneity in a LaAlO$_3$(001)/ (Pr$_{0.85}$Y$_{0.15}$)$_{0.7}$Ca$_{0.3}$CoO$_{3-\delta}$(9.5 nm) film by magnetic force microscopy (MFM).** (a-c) 5 K constant-height MFM images over the same nominal field of view, starting from a zero-field-cooled (ZFC) state. The out-of-plane applied magnetic field is (a) 0 kG, (b) 60 kG (6 T), and (c) returned to 0 kG. The tip height is 50 nm. For (a), the tip and sample biases were both 0 V, not adjusted to minimize electrostatic forces; the contribution from electrostatic forces was found negligible by comparing subsequent images with tip biases of 0 mV and 440 mV. For (b,c), the tip bias was 440 mV and the sample bias was 0 V. (d) Constant-height zero-field MFM image at 78 K (*i.e.*, above the Curie temperature (~60 K)); no magnetic contrast is distinguishable. The tip height here was again 50 nm, and the tip and sample biases were 400 mV and 0 V, respectively. (a-d) are on the same color scale (see right side of panel (d)) for Δ*f*, the shift of the cantilever resonant frequency. (e,f) 2D histograms of the values of Δ*f* at each pixel at 6 T (*y*-axis) *vs.* at 0 T

(*x*-axis) after ZFC (e) and after ramping down from 6 T (f). The histograms reveal only a weak correlation between ZFC and 6 T (e), but strong correlation (near-identical images) after ramping down the field. Prior to generating histograms, MFM images were aligned to correct for lateral drift between scans. (g) Dependence of $\Delta f$ on tip bias, with the sample grounded, at 5 different locations, with 50 nm tip height. The maximum represents the bias that minimizes the electrostatic force between sample and tip. There is no significant shift in the maxima, indicating that the features in MFM images are not electrostatic artifacts due to work function inhomogeneity. (h-k) Constant-lift 5 K MFM (right panels) and simultaneously acquired topographic images (left panels) at zero field after ramping down from 6 T (h,i), and at 6 T (j,k). As in (b,c), the MFM features are nearly identical in (i,k). The $15^{th}$- and $85^{th}$-percentile contours of the MFM images are superimposed on the topographic images to demonstrate that while some of the MFM features may be correlated with topography, the overall features in the MFM images are not purely due to topography. The tip lift was 30 nm and the tip and sample biases were 440 mV and 0 V, respectively. (i) and (k) have the same color scale for $\Delta f$. (h) and (j) have the same height scale.

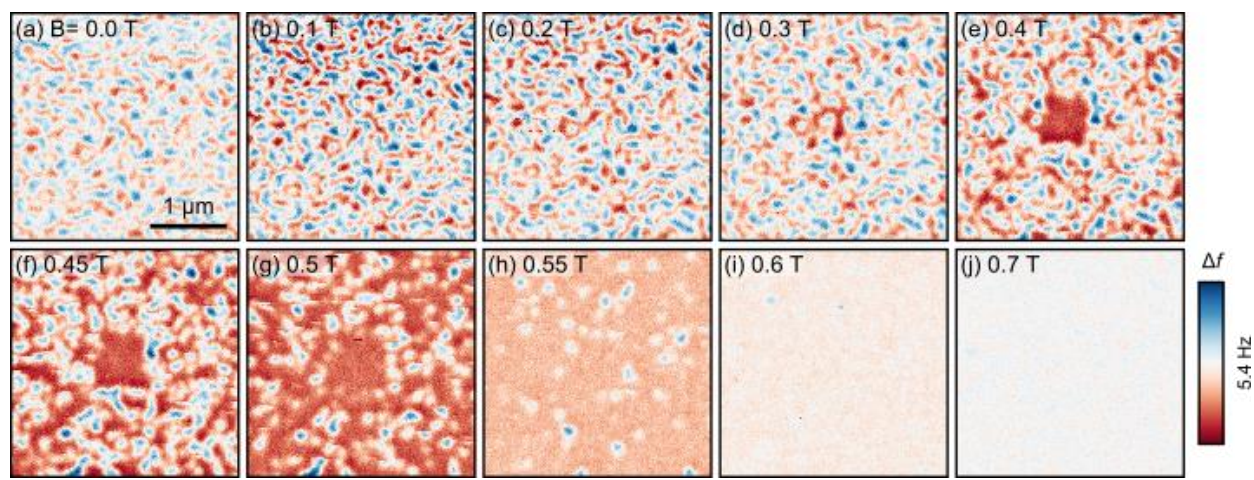

**Figure S8: Comparative characterization of magnetic domains in a ferromagnetic LaAlO$_3$(001)/La$_{0.5}$Sr$_{0.5}$CoO$_{3-\delta}$(13.9 nm) film by magnetic force microscopy (MFM).** 10 K constant-height MFM images over the same nominal field of view, starting from a zero-field-cooled (ZFC) state. The out-of-plane applied magnetic field is (a) 0, (b) 1 kG (0.1 T), (c) 2 kG (0.2 T), (d) 3 kG (0.3 T), (e) 4 kG (0.4 T), (f) 4.5 kG (0.45 T), (g) 5 kG (0.5 T), (h) 5.5 kG (0.55 T), (i) 6 kG (0.6 T), and (j) 7 kG (0.7 T). The tip height was 40 nm, and the tip and sample biases were 80 mV and 0 V, respectively. This film is a long-range-ordered phase-pure ferromagnet with out-of-plane anisotropy[5]. Consequently, panel (a) shows the labyrinth features expected from perpendicular ferromagnetic domains. Applying an out-of-plane field causes the domains aligned with the field (red) to grow, until, at 0.7 T, the entire field of view becomes a single domain. This is distinctly different to the behavior in Figs. S6 and S7.

Note: Between each MFM image, topographic scanning was done in a 600 nm square at the center of the field of view. During topographic scanning, the tip is close to the sample surface, causing the sample to experience a much larger stray field than during MFM imaging. As a result, the center of the MFM images in (c-h) became polarized at lower field than the rest of the field of view.

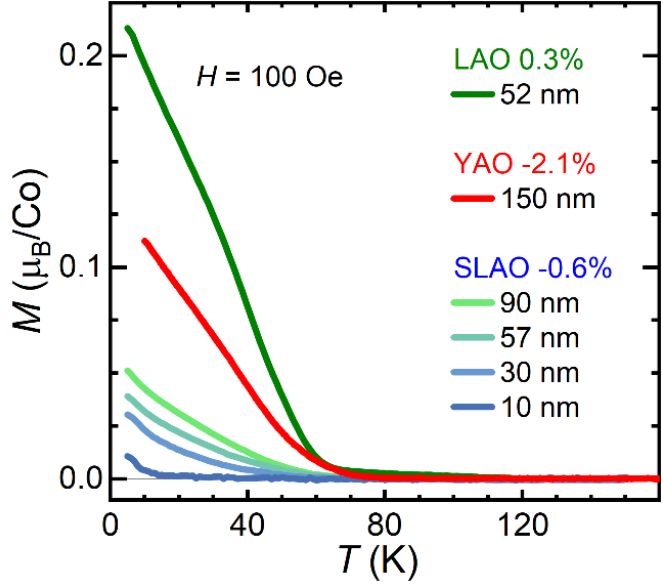

**Figure S9: Magnetic properties of a thick YAlO$_3$(101)/(Pr$_{0.85}$Y$_{0.15}$)$_{0.7}$Ca$_{0.3}$CoO$_{3-\delta}$ film.** Temperature ($T$) dependence of the magnetization ($M$) of a 150 nm film on YAlO$_3$(101) (red line). Measurements were made in an in-plane field of 100 G (10 mT) after field-cooling in 10 kG (1 T). For reference, a fully strained film on LAO (+0.3% in-plane strain, 52 nm thickness) (dark green line) is included, as well as 10 nm to 90 nm films on SLAO (-0.6% in-plane strain) (blue-green lines). Note that the magnitude of $M$ in the film on YAO should be treated as nominal, since the film will not be ferromagnetic through its depth[4] (unlike films on SLAO and LAO). As discussed in the main text, the Curie temperature of the 150 nm film on YAO is ~58 K.